\documentclass[usenatbib]{mn2e}

\usepackage[toc,page]{appendix}
\usepackage{amsmath}
\usepackage[dvips]{graphicx}
\usepackage{epsfig}
\usepackage{lscape}
\usepackage{natbib}
\usepackage{amssymb}
\usepackage{amsmath}
\usepackage{tabularx}
\usepackage{longtable}
\usepackage{float}

\def\rund#1{\left( #1 \right)}
\newcommand{\RE}{\theta_{\rm E}}
\newcommand{\Reff}{\theta_{\rm eff}}
\newcommand{\fdmbrein}{f_{\rm dm}(\leqslant\theta_{\rm E})}
\newcommand{\fdmlrein}{f_{\rm dm}(\theta_{\rm E})}
\newcommand{\fdmb}{f_{\rm dm}(\leqslant\theta)}
\newcommand{\fdml}{f_{\rm dm}(\theta)}
\newcommand{\Rfdmlhf}{\theta_{f50}}
\newcommand{\kms}{{\rm \,km\,s^{-1}}}

\newcommand{\CritDens}{\Sigma_{\rm cr}}
\newcommand{\Dd}{D_{\rm d}}
\newcommand{\Ds}{D_{\rm s}}
\newcommand{\Dds}{D_{\rm ds}}

\def\vc#1{{\mbox{\boldmath$#1$\unboldmath}}}

\title[Lens profiles, power-law models and bias on $H_0$] {Lens
  galaxies in the Illustris simulation: power-law models and the bias
  of the Hubble constant from time-delays} \author[Xu et al.] {Dandan
  Xu$^{1}$\thanks{E-mail: Dandan.Xu@h-its.org}, Dominique Sluse$^{2}$,
  Peter Schneider$^{3}$, Volker Springel$^{1,4}$, \and Mark
  Vogelsberger$^{5}$, Dylan Nelson$^{6}$, Lars Hernquist$^{6}$
  \\ $^{1}$ Heidelberg Institute for Theoretical Studies,
  Schloss-Wolfsbrunnenweg 35, 69118 Heidelberg, Germany \\ $^{2}$
  Institut d'Astrophysique et de G\'eophysique, Universit\'e de
  Li\`ege, All\'ee du 6 Ao\^ut 17, B5c, 4000 Li\`ege, Belgium
  \\ $^{3}$ Argelander-Institut f$\ddot{u}$r Astronomie,
  Universit$\ddot{a}$t Bonn, Auf dem H$\ddot{u}$gel 71, 53121 Bonn,
  Germany \\ $^{4}$ Zentrum f\"ur Astronomie der Universit\"at
  Heidelberg, Astronomisches Recheninstitut, M\"{o}nchhofstr. 12-14,
  69120 Heidelberg, Germany \\ $^{5}$ Department of Physics,
  Massachusetts Institute of Technology, 77 Massachusetts Avenue,
  Cambridge, United States \\ $^{6}$ Harvard Astronomy Department, 60
  Garden Street MS 46, Cambridge, MA 02138, United States }

\date{Accepted ...... Received ...... ; in original form......   }

\begin{document}
\pagerange{\pageref{firstpage}--\pageref{lastpage}} \pubyear{2015}
\maketitle
\label{firstpage}
\begin{abstract}

A power-law density model, i.e., $\rho(r) \propto r^{-\gamma'}$ has
been commonly employed in strong gravitational lensing studies,
including the so-called time-delay technique used to infer the Hubble
constant $H_0$. However, since the radial scale at which strong
lensing features are formed corresponds to the transition from the
dominance of baryonic matter to dark matter, there is no known reason
why galaxies should follow a power law in density. The assumption of a
power law artificially breaks the mass-sheet degeneracy, a well-known
invariance transformation in gravitational lensing which affects the
product of Hubble constant and time delay and can therefore cause a
bias in the determination of $H_0$ from the time-delay technique. In
this paper, we use the Illustris hydrodynamical simulations to
estimate the amplitude of this bias, and to understand how it is
related to observational properties of galaxies. Investigating a large
sample of Illustris galaxies that have velocity dispersion
$\sigma_{\rm SIE}\geqslant160\kms$ at redshifts below $z=1$, we find
that the bias on $H_0$ introduced by the power-law assumption can
reach $20\%-50\%$, with a scatter of $10\%-30\%$ (rms). However, we
find that by selecting galaxies with an inferred power-law model slope
close to isothermal, it is possible to reduce the bias on $H_0$ to
$\la5\%$, and the scatter to $\la10\%$. This could potentially be used
to form less biased statistical samples for $H_0$ measurements in the
upcoming large survey era.

\end{abstract}


\section{Introduction}
\label{sec:intro}

Strong gravitational lensing is a major tool of modern extragalactic
astrophysics. Ever since its first discovery \citep{Walsh1979Q0957},
it has been used as a natural telescope to magnify the distant
Universe, a scale to weigh galaxies, and a ladder to measure the
Hubble constant $H_0$ -- the expansion rate of the Universe
\citep{Refsdal1964TimedelayH0}. With the current and upcoming space
missions like GAIA and Euclid, and ground-based facilities such as the
Large Synoptic Survey Telescope (LSST) and the Square Kilometre Array,
strong lensing studies will experience an unprecedented opportunity to
exploit several thousands of lensing galaxies to be discovered (e.g.,
\citealt{Coe2009, Oguri2010, Finet2012, LSST2012}), while only
hundreds are currently known. This transition into the big data era
demands a good understanding of and control over systematic errors
that are introduced by lens modelling techniques, which will affect
scrutinized quantities such as the lens density slopes and their
evolution with time, and derived cosmological parameters such as
$H_0$.

The observational properties of galaxy-scale strong lens systems
(i.e. positions and flux ratios of unresolved images, brightness
distribution of lensed extended components, and time-delays) depend
mainly on the mass
distribution inside and near the Einstein radius of the lens, which corresponds
in general to a few half-light radii. Within this radius, both dark
and baryonic matter are believed to co-exist in roughly similar
amounts. However, because of the interplay between cooling and heating
mechanisms, the unknown efficiency of star formation, and a wide
variety of feedback processes, a large scatter in the galaxy total
mass profiles is
naively expected. Observationally, results from galaxy dynamics
combined with strong lensing measurements, but also X-ray emission of
elliptical galaxies, suggest a limited diversity of profiles and
indicate that the inner profile is almost isothermal (e.g.,
\citealt{Rusin2003, Humphrey2006, Gavazzi2007, Koopmans2009,
  Auger2010}). This surprising result, also known as the ``bulge-halo
conspiracy'', sets important constraints on the formation history of
galaxies (e.g., \citealt{Koopmans2006, Sonnenfeld2013b,
  Dutton2014}). In addition, it has an important impact on numerous
phenomenological studies of galaxies, as it motivates simple {\it
  power-law} models i.e., $\rho(r) \propto r^{-\gamma'}$ of the volume
mass density profile, with a corresponding convergence profile
$\kappa(\theta) \propto \theta^{1-\gamma^{\prime}}$ [i.e. the surface
density distribution $\Sigma({\bf \theta})$ normalized by the lensing
critical density 
\[
\CritDens=\left(\frac{c^2}{4\pi G}\right)
\left(\frac{\Ds}{\Dds\Dd}\right) \;, 
\]
where $\Dd$, $\Ds$ and $\Dds$ are
the angular diameter distances to the lens, to the source, and from
the lens to the source, respectively].
This model is commonly used in studies of galaxy-scale strong lens
systems, and often isothermality, i.e., $\gamma' = 2$, is assumed.

Despite observational support, the power-law density profile is no
more than a convenient ``working model'' for strong lensing
studies. This family of density profiles as found by fitting strong
lensing measurements {\it {does not uniquely/necessarily}} represent
the mass distribution of the inner regions of galaxies due to the
so-called mass-sheet degeneracy \citep[MSD;][]{Falco1985MST}. Any
given convergence field $\kappa({\vc \theta})$ can be transformed into
a new convergence profile $\kappa_{\lambda}({\vc \theta})$ via
\begin{equation}
\kappa_{\lambda}({\vc \theta})=\lambda\kappa({\vc \theta})+(1-\lambda).
\label{eq:MST}
\end{equation}
\noindent 
Together with an isotropic rescaling of the source plane $\beta \to
\lambda \beta$, the transformation (1) maintains the invariance of all
observable quantities, such as image positions and flux ratios, except
the time delays. This transformation is referred to as the
``mass-sheet transformation'' (MST). The only observable quantity
modified under the MST is the product of the Hubble constant $H_0$ and
the time delay $\Delta t$, which is transformed such that $H_0 \Delta
t \to \lambda H_0 \Delta t$. Because of the invariance of observables,
information from gravitational lens properties cannot distinguish
between any of the members of the family of mass profiles
$\kappa_\lambda(\vc \theta)$. However, if a particular parametrized
mass model is chosen, such as a (local or global) power-law profile,
the MST is artificially broken by yielding that member of the family
$\kappa_\lambda$ which is closest to the chosen mass profile
parametrization. This artificial breaking of the MSD can lead to a
systematic bias on the determination of $H_0$, for example, if the
logarithmic density profile near the Einstein radius is systematically
curved upwards or downwards relative to a power-law profile, then the
values of the transformation parameter $\lambda$ will be
systematically above or below unity \citep{SS13}.

The range of values of $\lambda$ expected from a population of lensing
galaxies is currently unknown. \citet{SS13} have presented an example
of a realistic galaxy model, composed as the sum of a baryonic and a
dark matter profile, which could be transformed into an approximate
power-law density profile through Eq. (\ref{eq:MST}) with $\lambda
\sim 1.2$, leading to a systematic error of about 20\% on $H_0$. Van
de Ven et al. (\citeyear{vandeven2009}) have shown that a large
variety of observationally motivated composite models of galaxies
systematically produce profiles that are slightly more convex/concave
than isothermal in regions probed by lensing. Here, we study the
expected distribution of surface density profiles of galaxies drawn
from a cosmological hydrodynamic simulation -- the Illustris Project
(\citealt{Illustris2014Nat}, see also \citealt{Illustris2014MN,
  Genel2014Illustris, Nelson2015IllustrisDataRelease}). By selecting
galaxies that most resemble the observed lensing galaxies in term of
mass/velocity dispersion, we test the validity of the power-law
assumption in the central regions of galaxies, and evaluate the
expected distribution of the multiplicative bias $\lambda$ on $H_0$
for this sample. We emphasize that in this analysis we assume that no
other external constraint than the lensing information is available to
break the MSD. Whereas stellar dynamical information from spectroscopy
can yield additional constraints on the lensing mass distribution and
thus helps to limit the allowed range of $\lambda$, the accuracy of
this method is limited, e.g., by spectroscopic resolution and the
unknown distribution of stellar orbits.
%

The outline of the paper is as follows. In Sect.\ 2, we describe the
methodology to measure the slope and curvature of the surface density
profiles of the simulated galaxies, and how we derive the $\lambda$
parameter which transforms the intrinsic distribution into an
approximate power law. In Sect.\ 3, we give a brief description of the
lensing galaxy samples that we have extracted from the Illustris
simulation; general properties of the simulated galaxy sample are
presented. As the (surface) density distributions are of particular
interest, we dedicate Sect.\ 4 to a detailed discussion about the
diversity of the central density profiles in regions probed by strong
lensing. In Sect.\ 5, we present distributions of $\lambda$ that
result from the use of power-law density profiles to model various
samples of lensing galaxies. Such distributions also indicate by how
much the derived value of $H_0$ would be systematically biased from
the true value by employing the power-law assumption. A final
discussion and conclusions are given in Sect.\ 6.

\section{Methodology}
\label{sec:method}

In this section we explain the quantities that we use to characterize
the density profiles of the simulated galaxies drawn from
Illustris. Since the MST only affects the radial density profile, but
leaves the shape of the isodensity contours invariant, we are only
concerned with the azimuthally-averaged, i.e., radial density profile
$\kappa(\theta)$ of a lensing galaxy.

In order to quantify how well power-law models can describe the
surface density profile of a lensing galaxy, we have first calculated
a mean local slope $s$ and the corresponding curvature $\xi$ following
the treatment of Schneider \& Sluse (2013). Assuming that the strong
lensing features (multiple compact images, images of extended source
components) are located between an inner and an outer radius of
$\theta_{1}=x_1\RE$ and $\theta_{2}=x_2\RE$, respectively, where $\RE$
is the Einstein radius of the lens, a mean logarithmic slope $s$ for
the convergence distribution $\kappa(\theta)$ between $\theta_{1}$ and
$\theta_{2}$ can then be calculated as\footnote{The quantity that we
  have denoted here as $s$ is actually the same as $\bar{s}$ of
  Schneider \& Sluse (2013).}
\begin{equation}
s\equiv\frac{\ln(\kappa_{2}/\kappa_{1})}{\ln(\theta_{1}/\theta_{2})},
\label{eq:sbar}
\end{equation}
where $\kappa_{1}\equiv\kappa(\theta_1)$ and
$\kappa_{2}\equiv\kappa(\theta_2)$ are the convergence at the inner
and outer image positions. $s=1$ corresponds to the projected slope
expected for an isothermal distribution; $s>1$ ($s<1$) indicates the
profile steeper (shallower) than isothermal. Unless mentioned
otherwise, we set $x_1=0.5$ and $x_2=1.5$ in this paper, which are
characteristic values for observed lensed systems.

We define the curvature parameter $\xi$ as
\begin{equation} 
\xi\equiv\frac{\kappa(\sqrt{\theta_{1}\theta_{2}})}
              {\sqrt{\kappa_{1}\kappa_{2}}},
\label{eq:xi}
\end{equation}
which indicates the closeness of the profile $\kappa(\theta)$ between
$\theta_{1}$ and $\theta_{2}$ to a power-law distribution: an exact
power-law profile will have $\xi=1$; while $\xi>1$ ($\xi<1$)
means that the projected density profile is concave-upward
(convex-downward), which lies above (below) the power-law
interpolation between $\theta_1$ and $\theta_2$.

The MST (1) transforms the original $\kappa(\theta)$ into a new
distribution $\kappa_{\lambda}(\theta)$, with slope
\[
s_{\lambda} =
\frac{\ln(\kappa_{\lambda}(\theta_2)/\kappa_{\lambda}(\theta_1))}
     {\ln(\theta_{1}/\theta_{2})}
\] 
and curvature
\[
\xi_{\lambda}=
     \frac{\kappa_\lambda(\sqrt{\theta_{1}\theta_{2}})}
          {\sqrt{\kappa_{\lambda}(\theta_1)
              \kappa_{\lambda}(\theta_2)}} \;, 
\]
leaving image astrometry and flux ratios unchanged. Here and below, we
specifically denote $\lambda$ as the value of the transformation
parameter that transforms 
$\kappa(\theta)$ to a new distribution $\kappa_{\lambda}(\theta)$,
which satisfies the criterion that $\xi_{\lambda}=1$; this condition yields
\begin{equation}
\lambda=\frac{\kappa_{2}+\kappa_{1}-2\xi\sqrt{\kappa_{\rm
      2}\kappa_{1}}}{\kappa_{2}+\kappa_{\rm
    1}-2\xi\sqrt{\kappa_{2}\kappa_{\rm
      1}}+(\xi^2-1)\kappa_{2}\kappa_{1}}.
\label{eq:lambda}
\end{equation}
Let $r\equiv\theta_1/\theta_2<1$, so that $\kappa_2/\kappa_1=r^s$
(from Eq.\ \ref{eq:sbar}). Then by dividing the numerator and
denominator of Eq. (\ref{eq:lambda}) by $\kappa_1$, we can re-write
$\lambda$ as a function of $r$, $s$, $\xi$ and $\kappa_2$:
\begin{equation}
\lambda={1+r^s-2\,\xi\, r^{s/2} \over 1+r^s-2\,\xi\, r^{s/2}
  +(\xi^2-1)\kappa_2}\;.
\label{eq:lambda0}
\end{equation}
Note that in the case where $\lambda=1$, $\xi=\xi_\lambda=1$. Here and
in the following, we use $\xi_\lambda=1$ as our power-law criterion:
that an MST with $\lambda$ given by Eq. (\ref{eq:lambda0}) transforms
$\kappa(\theta)$ into an approximate power law
$\kappa_\lambda(\theta)$ between $\theta_1$ and $\theta_2$ is in such
a sense that the three points $(\ln\theta_1,
~\ln\kappa_\lambda(\theta_1))$, $(\ln\theta_2,
~\ln\kappa_\lambda(\theta_2))$ and their mid-point
$(\ln\sqrt{\theta_1\theta_2}, ~\ln\kappa_\lambda
(\sqrt{\theta_1\theta_2}))$ lie on a straight line in the logarithmic
$\kappa(\theta)$ plot.


A meaningful MST requires that (i) the transformed density profile
$\kappa_{\lambda}(\theta)$ remains monotonically decreasing, which
yields $\lambda >0$; and (ii) $\kappa_{\lambda}(\theta)$ is
non-negative over the range considered, i.e.,
$\kappa_{\lambda}(\theta_2)>0$, which yields $\lambda<1/(1-\kappa_2)$.
Note that typically $85\%$ of the galaxies in our lens samples satisfy
condition (i) and all satisfy condition (ii). See Table 1 for the exact
fractions.

The slope $s_\lambda$ of the transformed profile, which by
construction is an approximate power law between $\theta_1$ and
$\theta_2$ (i.e., $\xi_\lambda=1$), can explicitly be related to the properties
of the original profile. Indeed, by definition it obeys the relation
\begin{equation}
  r^{s_{\lambda}}={\kappa_{\lambda}(\theta_2)\over \kappa_{\lambda}(\theta_1)}
  ={\lambda \kappa_2+(1-\lambda)\over \lambda \kappa_1+(1-\lambda)}
  ={\kappa_2+(1-\lambda)/\lambda \over \kappa_1+(1-\lambda)/\lambda}\;.
\label{eq:rs}
\end{equation}
From Eq. (\ref{eq:lambda}) we obtain
\begin{equation}
{1-\lambda\over\lambda}={(\xi^2-1)\kappa_1\kappa_2
\over \kappa_1+\kappa_2-2\xi\sqrt{\kappa_1 \kappa_2}}\;.
\label{eq:Lratio}
\end{equation}
Combining Eq. (\ref{eq:Lratio}) with Eq. (\ref{eq:rs}) then yields
\begin{equation}    
r^{s_{\lambda}} = {\kappa_2^2-2\xi\kappa_2\sqrt{\kappa_1
      \kappa_2} + \xi^2\kappa_1\kappa_2
\over
\kappa_1^2-2\xi\kappa_1\sqrt{\kappa_1
      \kappa_2} + \xi^2\kappa_1\kappa_2} \;.
\label{eq:rslambda}  
\end{equation}       
By dividing the numerator and denominator of Eq. (\ref{eq:rslambda}) by
$\kappa_1\kappa_2$ we obtain
\begin{eqnarray}             
r^{s_{\lambda}}&=& {r^{s}-2\xi r^{s/2} + \xi^2 
\over r^{-s}-2\xi r^{-s/2} + \xi^2 } \nonumber \\   
&=& r^{s} \rund{\xi - r^{s/2} \over
1 - \xi \, r^{s/2}  }^2 \\
&=& r^{s} \rund{ 1+ \frac{\left(r^{-s}-1\right)
  \left(\xi^2-1\right)} {\left(r^{-s/2} -\xi\right)^2} }
\;. \nonumber
\end{eqnarray}
Hence, the power-law slope after MST is
\begin{equation}
s_{\lambda}=s + \ln \left(1+ \frac{\left(r^{-s}-1\right)
  \left(\xi^2-1\right)} {\left(r^{-s/2} -\xi\right)^2} \right) /\ln r \;.
\label{eq:slambda}
\end{equation}
Thus, $s_{\lambda}$ is a function solely of the original slope $s$ and
the curvature parameter $\xi$. If $\xi>1$, then (since $r^{-s}>1$ and
$\ln r<0$) $s_{\lambda} < s$, i.e., in this case the transformed
profile is flatter than the original. This case also corresponds to
$\lambda <1$, i.e., the mass sheet added has positive convergence,
leading to a flatter profile after MST. 

It is worth noting that a different radial range
$[\theta_1,~\theta_2]$ would yield different $\kappa_{\lambda}$ and
$\lambda$. This is simply because they depend on the slope $s$ and
curvature $\xi$ which are defined in terms of the angular range where
strong lensing measurements are available. As stated before, we focus
on the range $\theta_1=0.5\,\RE$ and $\theta_2=1.5\,\RE$. Results for
a different angular range (with $\theta_1=0.8\,\RE$ and
$\theta_2=1.2\,\RE$) are also presented in
Appendix~\ref{AppendixA}. As can be seen, although measurements in
these two cases are not exactly the same on a one-to-one basis, the
statistical distributions of $\lambda$ are independent of the choice
of the $\theta_i$ ($i=1,~2$).

Equations (1) to (10) relate to the local convergence profile
$\kappa(\theta)$. They were derived and discussed in the framework of
the assumption that the convergence profile follows approximately a
power law in the range $\theta_1\leqslant \theta\leqslant
\theta_2$. Alternatively, we can also define analogous quantities
under the assumption that the radial dependence of the deflection
angle behaves like an approximate power law, which is equivalent to a
power law in the mean convergence within $\theta$, i.e.,
\[
\bar\kappa(\theta)={2\over \theta^2}\int_0^\theta
{\rm d}\theta' \;\theta'\,\kappa(\theta')\;.
\]
Hence, we can consider an MST that results in a transformed cumulative
distribution $\bar{\kappa}_{\lambda}$ to satisfy our power-law
criterion so that the profile of $\bar{\kappa}_{\lambda}$ is a good
approximation of a power law between $\theta_1$ and $\theta_2$.  The
MST for $\bar\kappa$ is the same as that for $\kappa$, i.e.,
\begin{equation}
\bar{\kappa}_{\lambda}={\lambda}\bar{\kappa}+(1-{\lambda}).
\label{eq:MSTofKbar}
\end{equation}
Note that the radii where $\kappa=\kappa_{\lambda}=1$ and where
$\bar{\kappa}=\bar{\kappa}_{\lambda}=1$ remain unchanged; the latter
corresponds to the Einstein radius by definition. Hereafter we denote
by barred symbols all quantities that are associated with the
cumulative distribution $\bar{\kappa}$ and its MST. The whole formalism
from Eq. (2) to (10) also holds for $\bar{\kappa}$ after replacing
$\kappa$ with $\bar{\kappa}$, $s$ and $\xi$ with $\bar{s}$ and
$\bar{\xi}$, $s_{\lambda}$ and $\xi_{\lambda}$ with
$\bar{s}_{\lambda}$ and $\bar{\xi}_{\lambda}$, and $\lambda$ with
$\bar{\lambda}$.

Which of the two MSTs -- the one that yields an approximate local
power law of the convergence, i.e., $\xi_\lambda=1$, or that leading
to an approximate power law in the deflection, i.e.,
$\bar\xi_\lambda=1$ -- is the more relevant one depends on which
assumptions are made in lens modelling. In many applications, it is
assumed that the density profile follows a global power law (at least
up to the radius where strong lensing features are observed), in which
case the requirement $\bar\xi_\lambda=1$ applies. Alternatively, lens
models can be considered in which the local profile follows a power
law, but an additional mass component must be assumed on angular
scales smaller than $\theta_1$, i.e., the presence of a supermassive
black hole, or a finite core radius of the mass distribution, both of
which cause deviations from a global power law. Note that the
assumption of a global power law implies that the density profile
between $\theta_1$ and $\theta_2$ is a local power law (with the same
slope). Thus, the global power-law assumption is a stronger one than
just enforcing a local power law. However, since the density profiles
of real galaxies do not follow power laws, this conclusion does not
hold for real lenses.  We note that in general, $\lambda \neq
\bar{\lambda}$, and $s_{\lambda} \neq \bar{s}_{\lambda}$.

\section{The Illustris lens samples and their general properties}
\label{sec:Illustris}
The Illustris Project is a series of cosmological hydrodynamical
simulations of galaxy formation (\citealt{Illustris2014Nat}, see also
\citealt{Illustris2014MN, Genel2014Illustris,
  Nelson2015IllustrisDataRelease}). The highest resolution run covers
a volume of (106.5 Mpc)$^3$ and has a dark matter mass resolution of
$6.26\times10^6M_{\odot}$ and an initial baryonic mass resolution of
$1.26\times10^6M_{\odot}$, resolving gravitational dynamics down to a
physical scale of $\epsilon=710$ pc. Taking into account various
baryonic processes (\citealt{Vogelsberger2013pm,
  Vogelsberger2014err}), such as gas cooling, stellar evolution and
feedback, chemical enrichment, supermassive black hole growth and
feedback from active galactic nuclei, the Illustris simulation
resolves 40\,000 galaxies with a variety of morphologies and
reproduces many fundamental properties of observed galaxies (e.g.,
galaxy luminosity functions and Tully-Fisher relations).

The Illustris Project is an ideal ``laboratory'' for our study. On the
one hand, this is because galaxy-scale strong lensing probes the
projected central regions of galaxies where baryons yield a
significant contribution to the total mass profile. On the other hand,
the Illustris simulation provides a large and realistic sample of
galaxies as needed to quantify in a statistical way the systematics
associated with the assumption of power law density profiles in
strong-lensing studies.

\subsection{Main samples of lensing galaxies}

We take the lensing galaxies at various redshifts from the mock
strong-lens catalogue of the Illustris simulation
(\citealt{Nelson2015IllustrisDataRelease}, see also Xu et al., in
preparation). Our selection criteria are as follows: for a given
source redshift $z_{\rm s}$, the angular (and physical) Einstein
radius $\RE$ (and $R_{\rm E}$) of each galaxy is first determined as
the radius within which the mean convergence
$\bar{\kappa}=1$. Referring to the singular isothermal sphere model,
we characterize each lens galaxy by its ``velocity dispersion''
$\sigma_{\rm SIE}$, which is linked to the Einstein radius via
$\RE=4\pi(\sigma_{\rm SIE}/c)^2 ({\Dds}/{\Ds})$. We select galaxies
that have $\sigma_{\rm SIE}\geqslant 160\kms$. This lower limit on the
velocity dispersion is motivated by the observed lens sample from the
SLACS survey, where stellar velocity dispersions of lens galaxies
range from $160\kms$ to $400\kms$ \citep{Bolton2008SLACSV}.

The Illustris simulation does not sample the very massive end of the
mass spectrum well, i.e., galaxy clusters. There are a couple of
massive systems that have total masses around
$10^{14}h^{-1}M_{\odot}$. We have adopted a further selection
criterion, i.e., excluding all satellite/companion galaxies in the
group and galaxy environment and only selecting the ``central''
galaxies to form our sample. The projection effects from their
companions (as well as galaxies along the line of sight) have been
excluded from the calculation of the density profiles of these
``central'' galaxies.

Note that the surface brightness distributions of the Illustris
lensing galaxies can be well fitted by Sersic profiles
(\citealt{Sersic1963}). In particular for the elliptical
galaxies\footnote{The surface brightness distribution of each galaxy
  in the mock catalogue has been fitted using both the de Vaucouleurs
  and the exponential disc profiles. A galaxy is classified as an
  early (late) type if the former (latter) provides a better fit. }
whose stellar masses and velocity dispersions are in the same ranges
as the SLACS samples, their logarithmic density slopes are also
consistent with the one constrained for the SLACS samples (e.g.,
\citealt{Auger2010, Sonnenfeld2013b}). A detailed comparison of
Illustris galaxies to the SLACS sample will be presented in a
forthcoming paper (Xu et al., in preparation). In this paper, as we
are interested in the {\it surface} density distribution, which is a
projected quantity, and in order to increase the sample size, we treat
the three independent projections of each selected galaxy as
independent lenses. This yields more than 1000 surface density
profiles for the analysis at each of the lens redshifts studied.

\begin{table*} 
\caption{A summary of sample properties (galaxies in their three
  independent projections with $\sigma_{\rm SIE}\geqslant 160\kms$ are
  selected). The total number of galaxy projections that meet the
  selection criterion are given in row (3); row (4), (5) and (6) give
  the percentages of galaxies whose profiles can be transformed to
  power laws via MST (see Sect. 2); $R_{\rm E}^{\rm min}$ in row (7)
  is the minimal physical Einstein radius, which corresponds to the
  lower limit of $\sigma_{\rm SIE}=160\kms$; in row (8), (9) and (10),
  the mean and medi\ an Einstein radius and its standard deviation
  $\sigma_{R_{\rm E}}$ are presented, respectively.}
\begin{minipage}{\textwidth}
\begin{tabular}{c c c c c | c c c} \hline\hline
Sample sets & \multicolumn{4}{c|}{$z_{\rm s}=1.5$} &
\multicolumn{3}{c}{$z_{\rm d}=0.6$} \\ \hline Redshifts & ~~$z_{\rm
  d}=0.2$~~ & ~~$z_{\rm d}=0.4$~~ & ~~$z_{\rm d}=0.6$~~ & ~~$z_{\rm
  d}=0.8$~~ & ~~$z_{\rm s}=1.0$~~ & ~~$z_{\rm s}=1.5$~~ & ~~$z_{\rm
  s}=3.0$~~ \\\hline Total number of projections & 1044 & 1334 & 1433
& 1363 & 1092 & 1433 & 1673 \\ Meaningful MST for $\kappa$ & 90\% &
93\% & 90\% & 88\% & 88\% & 90\% & 93\% \\ Meaningful MST for
$\bar{\kappa}$ & 77\% & 82\% & 78\% & 75\% & 73\% & 78\% & 81\%
\\ Meaningful MST for both $\kappa$ and $\bar{\kappa}$ & 76\% & 81\% &
77\% & 73\% & 72\% & 77\% & 80\% \\ $R_{\rm E}^{\rm min}$ (kpc) & 2.00
& 2.60 & 2.50 & 2.06 & 1.68 & 2.50 & 3.29 \\ Mean $R_{\rm E}$ (kpc) &
4.16 & 5.32 & 5.14 & 4.41 & 3.71 & 5.14 & 6.47 \\ Median $R_{\rm E}$
(kpc) & 3.62 & 4.69 & 4.49 & 3.91 & 3.29 & 4.49 & 5.60 \\ Standard
deviation $\sigma_{R_{\rm E}}$ (kpc) & 2.04 & 2.53 & 2.46 & 2.04 &
1.75 & 2.46 & 3.15 \\ \hline\hline
\end{tabular}
\end{minipage}
\begin{flushleft}
\end{flushleft}
\label{tab:TableSample}
\end{table*}

In the following we refer to the region where strong lensing
measurements are available (i.e., the range $[\theta_1, \theta_2]$) as
{\it the strong lensing region}. Because $\theta_{\rm E}$ is an
angular quantity, the strong lensing region depends on the lens
redshift $z_{\rm d}$ and the source redshift $z_{\rm s}$. In order to
separate the effect of galaxy evolution from a change of $R_{\rm E}$
(such that lensing probes different regions of a same galaxy), we have
formed two sets of samples according to different combinations of
$z_{\rm s}$ and $z_{\rm d}$.

In the first set, we have selected galaxies at $z_{\rm
  d}=[0.2,~0.4,~0.6,~0.8]$, assuming a fixed source redshift at
$z_{\rm s}=1.5$ (Table \ref{tab:TableSample}, left-hand columns). Note
that assuming the innermost boundary of the strong lensing region to
be $0.5R_{\rm E}$, the $z_{\rm d}-z_{\rm s}$ combinations used here
guarantee that even the smallest galaxies (with Einstein radius of
$R_{\rm E}^{\rm min}$, which corresponds to the lower limit of
$\sigma_{\rm SIE}=160\kms$) are fully resolved at radii larger than
the innermost boundary, i.e., $0.5R_{\rm E}^{\rm min}\ga \epsilon$
(where $\epsilon=710$ pc is the simulation softening length). As can
be seen from Table \ref{tab:TableSample}, the mean Einstein radius of
the galaxies, $\left<R_{\rm E}\right>$, is approximately the same for
the four $z_{\rm d}-z_{\rm s}$ combinations. We therefore probe with
this sample typically the inner 5\,kpc of galaxies, such that any
difference in the final statistical results is due to galaxy
evolution.

In the second sample set, we fix the lens redshift at $z_{\rm d}=0.6$
and assume source redshifts at $z_{\rm s}=[1.0,~1.5,~3.0]$
(Table~\ref{tab:TableSample}, right-hand columns). Again the $z_{\rm
  d}-z_{\rm s}$ combinations applied in this sample set also guarantee
that the smallest galaxies are fully resolved at radii larger than the
innermost boundary of their strong lensing regions. For fixed $z_{\rm
  d}$, the higher $R_{\rm E}$ correspond to the larger $z_{\rm s}$:
the mean $\left<R_{\rm E}\right>$ for the $z_{\rm s}=3.0$ sample is
nearly twice as large as that for the $z_{\rm s}=1.0$ sample. This
allows us to study the dependence of our results on the radii probed
by strong lensing. Note that different numbers of galaxies are
obtained in spite of the same $\sigma_{\rm SIE}$ cut, because the
actual galaxy mass distributions are not truly isothermal, which thus
introduces scatter in the $\sigma_{\rm SIE}$-mass relation. The
scatter in mass that corresponds to $\sigma_{\rm SIE}=160\kms$ for the
$z_{\rm s}=1.0$ sample results in a minimum halo mass of $\sim
9\times10^{11}h^{-1}M_{\odot}$; while the scatter in the $z_{\rm
  s}=3.0$ sample leads to a slightly lower halo-mass limit of $\sim
7\times10^{11}h^{-1}M_{\odot}$. The latter sample therefore has a
larger number of galaxies than the former.

\begin{figure}
\centering
\includegraphics[width=8.4cm]{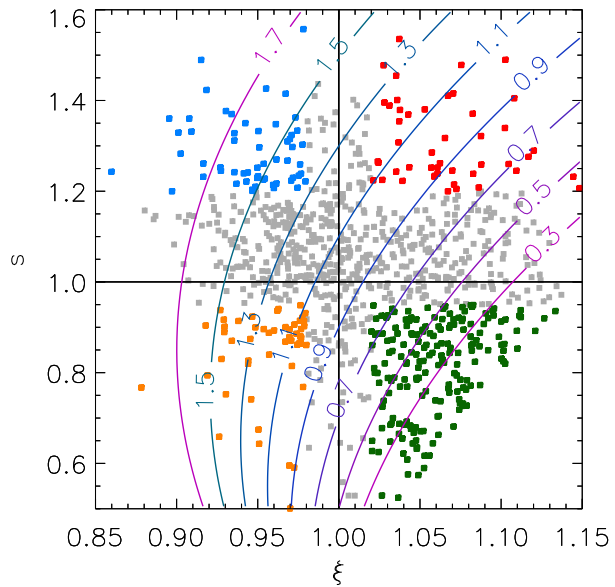}
\caption{The $s-\xi$ distribution of the selected lensing galaxies at
  $z_{\rm d}=0.2$, assuming $z_{\rm s}=1.5$. The sample is divided
  into four subgroups according to their $s$ and $\xi$ values: subgroup I
  (red) is defined as galaxies with $s>1.2$, $\xi>1.02$; subgroup II
  (blue) are those with $s>1.2$, $\xi<0.98$; subgroup III (orange)
  represents galaxies with $s<0.95$, $\xi<0.98$; and subgroup IV
  (green) are those with $s<0.95$, $\xi>1.02$. Contours indicate where
  the transformed slopes $s_{\lambda} =
  [0.3,~0.5,~0.7,~0.9,~1.1,~1.3,~1.5,~1.7]$. }
\label{fig:GeneralPropt0}
\end{figure}

\subsection{Galaxy profiles and general properties}
\label{sec:subgroups}

In the following, we present the surface density profiles of the
selected galaxies and some of their general properties. For each
galaxy, we have azimuthally averaged\footnote{We have also calculated
  the radial profile by averaging the convergence distribution in
  elliptical annuli, the shape of which is determined by the second
  moment of total projected mass within 2$\RE$. We have verified that
  both the curvature parameter $\xi$ and the MST parameters $\lambda$
  under the elliptical average are consistent with the ones derived
  under azimuthal averaging; the final result remains. } the surface
mass density distribution to obtain the radial profile of the
convergence in the projected central region. To derive the profile
parameters, we fit a 10-order polynomial to the logarithmic radial
profile within a radius range from $2\epsilon$ to five times the
half-stellar-mass radius of the galaxy, which well covers the region
constrained by strong lensing. Using the polynomial fitted radial
profiles, we measure the slopes and curvatures of both the local
convergence $\kappa$ and the cumulative distribution $\bar{\kappa}$,
which are then used to derive the MST parameters $\lambda$ and
$\bar{\lambda}$ and the transformed slopes $s_\lambda$ and
$\bar{s}_\lambda$, according to the formalisms derived in Sect.\ 2.

Figure \ref{fig:GeneralPropt0} shows the distribution of $s$ versus
$\xi$ for the lens sample with $z_{\rm d}=0.2$ and $z_{\rm
  s}=1.5$. The overlaid contours indicate the transformed slope
$s_{\lambda}$ as a function of ($s$, $\xi$), which increases from the
bottom right quadrant to the top-left quadrant in the $s-\xi$
plane. The mean slope $s$ measured between 0.5$\,\RE$ and 1.5$\,\RE$
typically ranges from 0.5 to 1.6, while for the curvature $\xi$, we
have $|\xi-1|\la 15\%$ for all our samples.  Of course, the range of
$\xi$ strongly depends on $\Delta\theta\equiv\theta_2-\theta_1$: as
$\Delta\theta\rightarrow 0$, $\xi\rightarrow 1$. Note that galaxies
that fail the criteria ``$\lambda>0$'' (and thus $s_{\lambda}>0$) for
a meaningful MST lie at the lower-right corner of the diagram, to the
right of the green points. These galaxies are not fundamentally
different from the ``green'' galaxies in terms of their physical
properties except that they have even shallower profiles with larger
curvatures. Mathematically in order to transform their profiles into
power laws, the additional mass sheet (1-$\lambda$) has to be so large
that the transformed power-law densities become lower in the centres
and larger at the outskirts, and thus $s_\lambda$ becomes negative.

In order to understand different profile behaviours, we have further
divided each lens sample into four representative subgroups according
to their $s$ and $\xi$ values. As shown by the coloured symbols in
Fig. \ref{fig:GeneralPropt0}, subgroup I as indicated by red is
defined for galaxies with $s>1.2$ and $\xi>1.02$, i.e., galaxies with
mean slopes steeper than isothermal and having concave-upward
profiles; subgroup II as shown in blue consists of galaxies with
$s>1.2$ and $\xi<0.98$, i.e., galaxies that also have mean slopes
steeper than isothermal but convex-downward profiles; subgroup III as
marked in orange represents galaxies with $s<0.95$, $\xi<0.98$, i.e.,
those with mean slopes flatter than isothermal and with
convex-downward profiles; and subgroup IV plotted in green is composed
of galaxies with $s<0.95$, $\xi>1.02$, i.e., those with mean slopes
flatter than isothermal but having concave-upward profiles.

Since $\xi=1$ indicates approximate power-law profiles, it is used to
separate subgroups. The division using $s=1$ is somewhat arbitrary;
this is purely motivated by the observation that the distribution of
the total density slopes of lens galaxies seems to peak at the
isothermal slope (\citealt{Rusin2003, Rusin2005, Koopmans2006,
  Koopmans2009, Gavazzi2007, Auger2010}). In other words, the four
subgroups are extreme cases where the density profiles are far from
isothermal and perfect power law.

Apart from the profile parameters, for each galaxy in projection we
have also calculated a variety of properties:
(1) the total stellar mass $M_{\star}$ of the galaxy and the total
mass $M_{\rm tot}$ of the galaxy plus its host dark matter halo; (2)
the effective radius $\Reff$, within which half of the galaxy
luminosity is enclosed; (3) the cumulative dark matter fraction
$\fdmbrein$ projected {\it within} $\RE$ and the local dark matter
fraction $\fdmlrein$ projected {\it at} $\RE$; (4) the angular radius
$\Rfdmlhf$ where the projected density distribution of dark matter
intercepts that of the baryonic matter, i.e., the radius where the
local dark matter fraction $f_{\rm dm}(\Rfdmlhf)=0.5$.

\begin{figure}
\centering
\includegraphics[width=8cm]{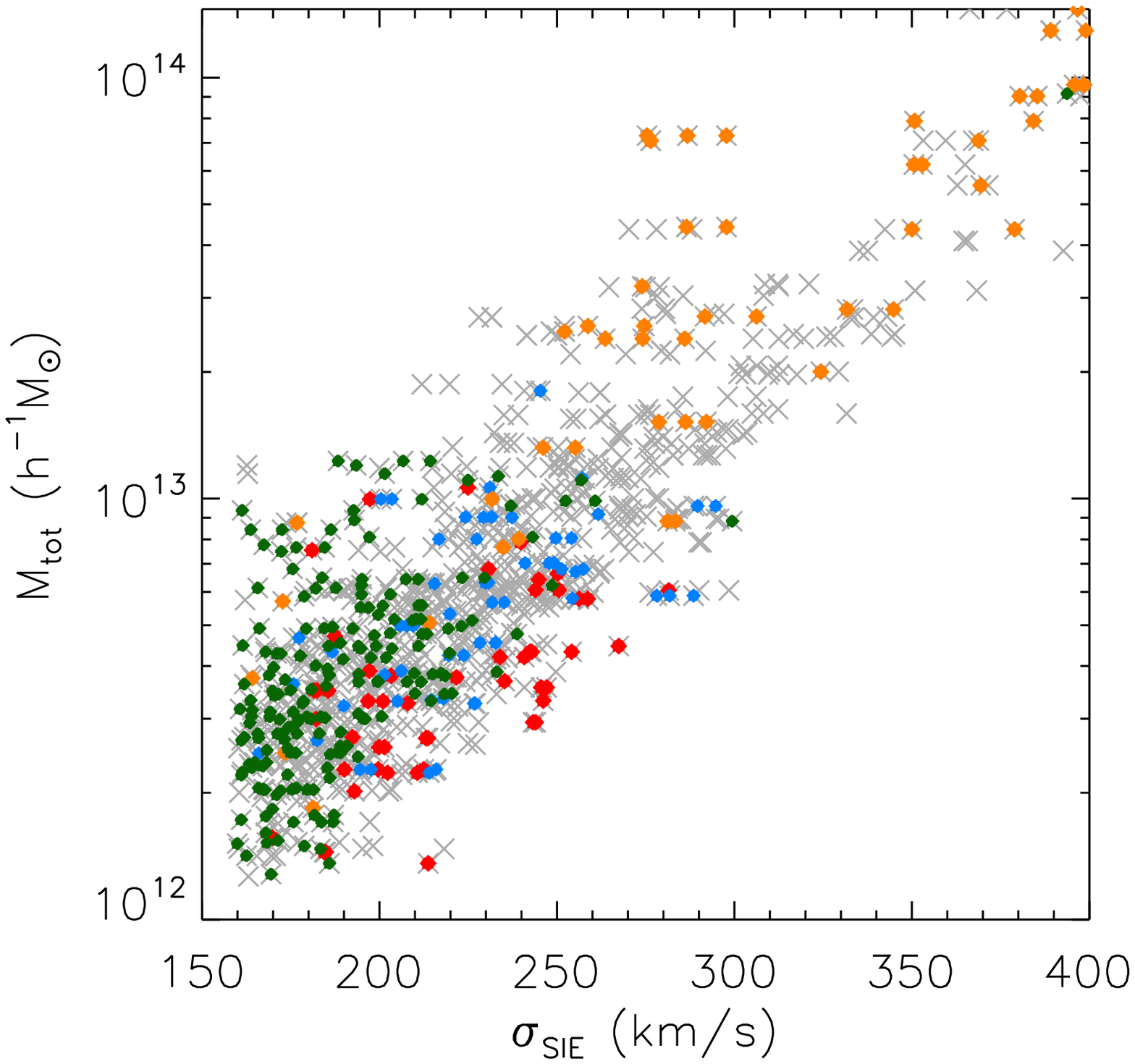}
\includegraphics[width=8cm]{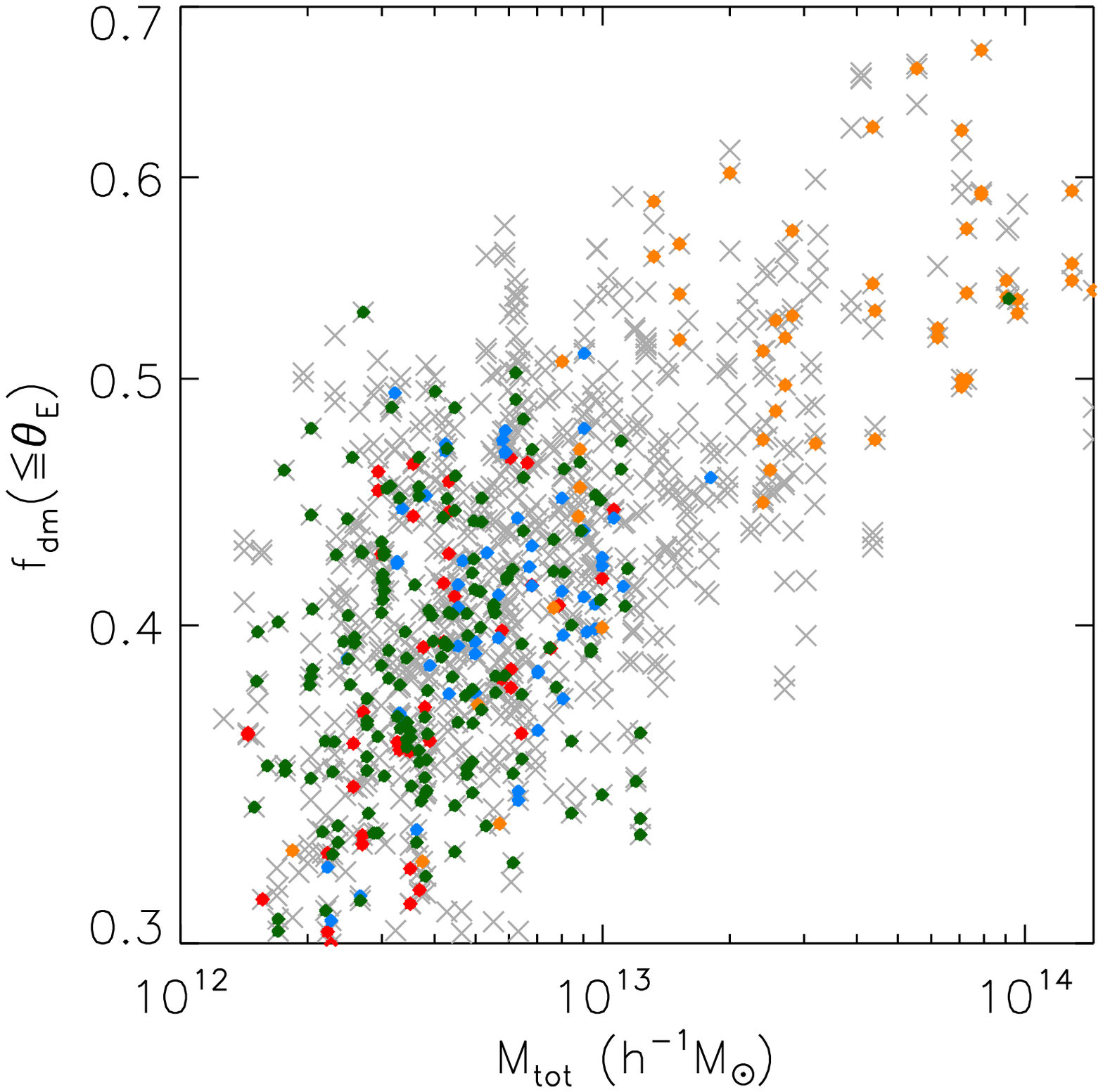}
\caption{The distributions of $M_{\rm tot}$ versus $\sigma_{\rm SIE}$
  (top) and $\fdmbrein$ versus $M_{\rm tot}$ (bottom). The galaxy
  sample as well as the colour coding is the same as used in
  Fig. \ref{fig:GeneralPropt0}.  }
\label{fig:GeneralPropt1}
\end{figure}

Fig. \ref{fig:GeneralPropt1} shows the distributions of $M_{\rm tot}$
versus $\sigma_{\rm SIE}$ (top) and $\fdmbrein$ versus $M_{\rm tot}$
(bottom) for the same galaxy sample as in
Fig. \ref{fig:GeneralPropt0}. Different projections of the same galaxy
result in multiple values of $\sigma_{\rm SIE}$ and $\fdmbrein$ at the
same mass, reflecting the triaxiality of the lens mass
distribution. The cumulative dark matter fraction $\fdmbrein$
typically ranges from 30\% to 70\% with more massive galaxies having
higher $\fdmbrein$, in good agreement with strong lensing observations
(e.g., \citealt{TK2004}; Koopmans et al. 2006; \citealt{Jiang2007,
  Tortora2009, Grillo2009, Cardone2009, Napolitano2010}).

It can be seen clearly that the four subgroups have different galaxy
properties, e.g., the shallower- and convex-profiled galaxies
(subgroup III) are among the most massive ones and have the largest
velocity dispersions and the highest dark matter fractions. In the
next section, we aim at understanding the direct reason for such a
diversity in the surface density profiles measured in strong lensing
regions.

\section{The central surface density profile: slope and curvature}
\label{sec:slope}

Galaxy-scale strong lensing probes the central few kpc region of a
lensing galaxy. The central (surface) density profile is of particular
interest, because on such scales, the enclosed mass of dark and
baryonic matter are quite similar. Various processes associated with
star formation and feedback together shape the density distributions
of both components and their sum.

\begin{figure*}
\centering
\includegraphics[width=16cm]{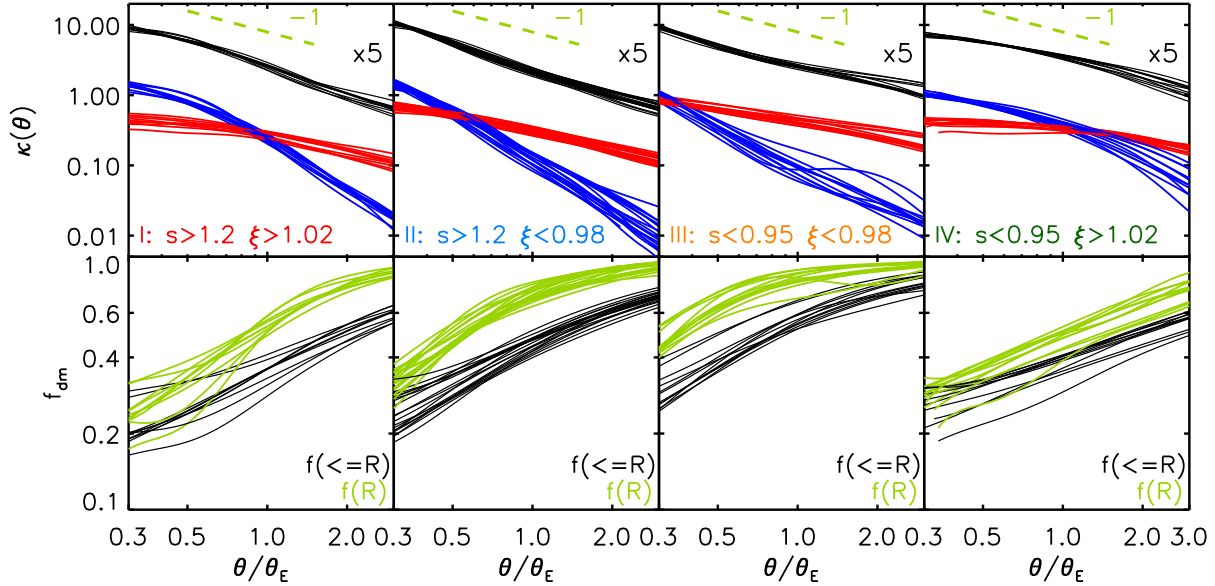}
\caption{From each subgroup of the lens sample used in
  Fig. \ref{fig:GeneralPropt0}, ten typical surface density profiles
  and projected dark matter fraction distributions are shown. In the
  top panels the black curves show the total surface density
  distribution $\kappa(\theta)$, scaled up by a factor of 5 for
  clarity; the red and blue curves represent profiles of the projected
  dark matter $\kappa(\theta) f_{\rm dm}(\theta)$ and projected
  baryonic mass $\kappa(\theta) [1-f_{\rm dm}(\theta)]$, respectively;
  the dashed green line indicates the logarithmic slope of $s=1$ from
  0.5$\theta_{\rm E}$ to 1.5$\theta_{\rm E}$. In the bottom panels,
  the distributions of the cumulative and local dark matter fraction
  $f_{\rm dm}(\leqslant \theta)$ and $f_{\rm dm}(\theta)$ are given in
  black and green, respectively. }
\label{fig:DensityProfile}
\end{figure*}

We first illustrate in Fig. \ref{fig:DensityProfile} the projected
density profiles associated with the four galaxy subgroups
(Sect.~\ref{sec:subgroups}). For each subgroup we show ten typical
density profiles and their dark matter fraction distributions. The
projected radii are normalized to $\RE$ (see
Fig.~\ref{fig:DensityProfile2Reff} for profiles as a function of
$\Reff$). In the top panels, the black curves are the total surface
density distributions $\kappa(\theta)$, scaled up by a factor of 5 for
clarity; the red and blue curves represent profiles of the projected
dark matter $\kappa(\theta) f_{\rm dm}(\theta)$ and projected baryonic
mass $\kappa(\theta) [1-f_{\rm dm}(\theta)]$, respectively; the dashed
green line indicates the logarithmic slope of $s=1$ from
0.5\,$\theta_{\rm E}$ to 1.5\,$\theta_{\rm E}$. In the bottom panels,
the cumulative and local dark matter fraction $\fdmb$ and $\fdml$ are
given in black and green, respectively.

The first conclusion is that, as expected, the baryonic
distribution is much more concentrated than the dark matter
distribution. In regions probed by strong lensing, the two profiles
intercept; the slope of the former is in general steeper than that of
the latter. As a result, baryons and dark matter dominate different
parts of the total (surface) density profile: the former in the inner
regions and the latter at larger radii; the dark matter fraction
therefore increases with radius. 


\subsection{Individual profiles of the two ingredients}

As the total density profile is the sum of two components, we present
hereafter slopes $s$ and curvatures $\xi$ of the baryonic
(i.e. $s_{\rm b}$ and $\xi_{\rm b}$) and dark matter profile
(i.e. $s_{\rm dm}$ and $\xi_{\rm dm}$) separately.

\begin{figure}
\centering
\includegraphics[width=8.5cm]{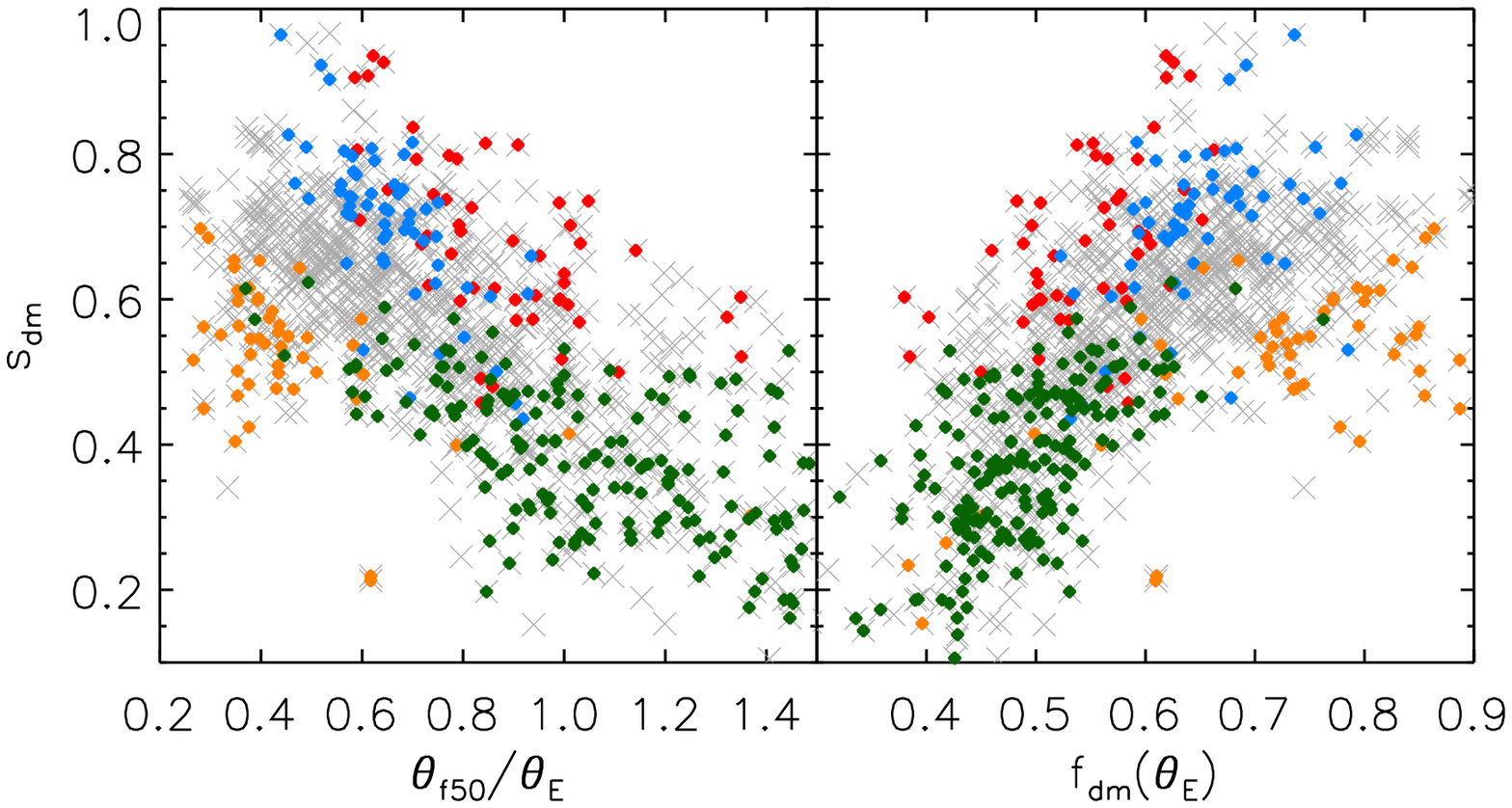}
\includegraphics[width=8.5cm]{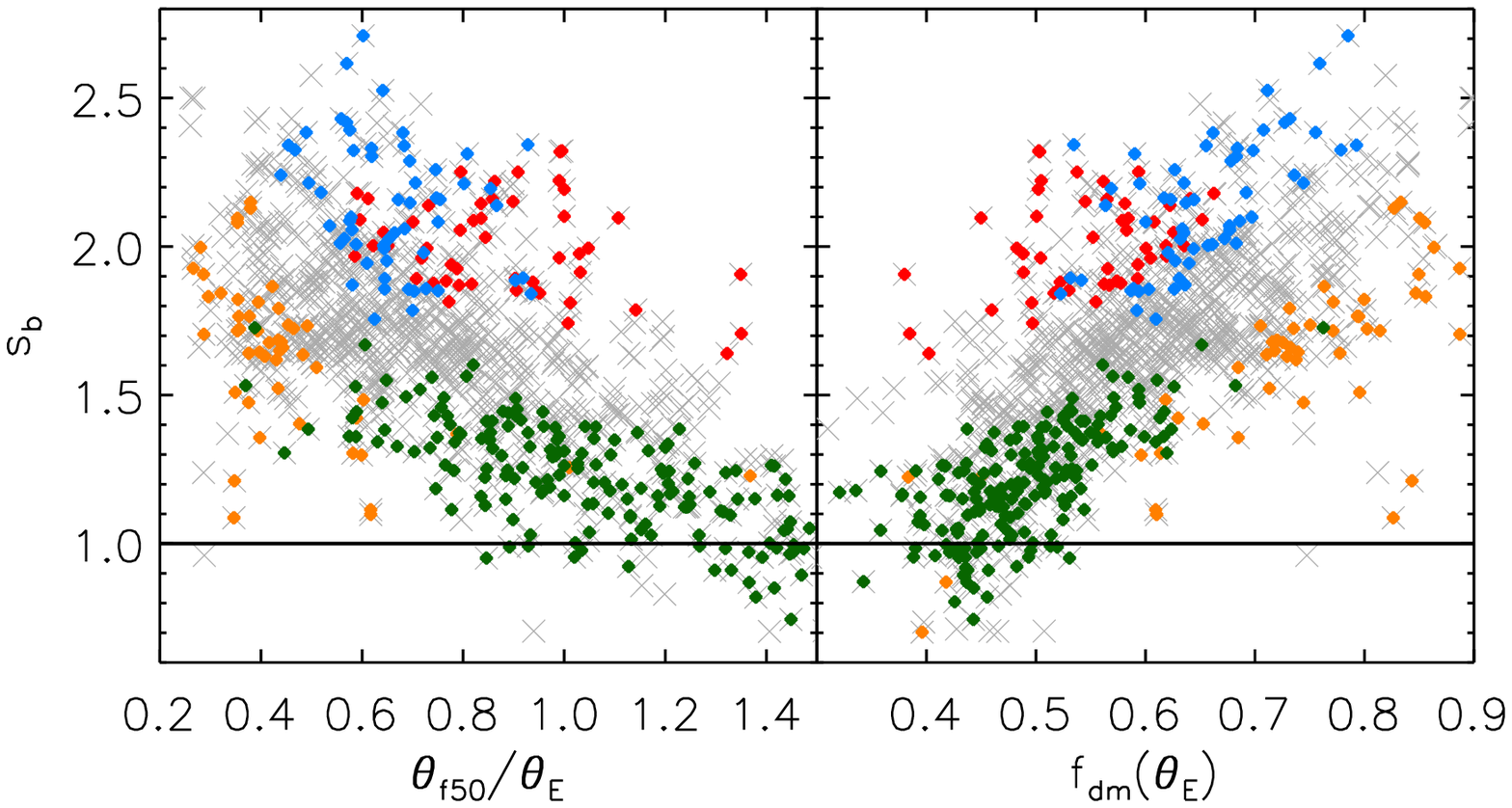}
\caption{Top left: dark matter slope $s_{\rm dm}$ versus
  $\Rfdmlhf/\RE$. Top right: $s_{\rm dm}$ versus $\fdmlrein$.  Bottom
  left: baryonic slope $s_{\rm b}$ versus $\Rfdmlhf/\RE$. Bottom
  right: $s_{\rm b}$ versus $\fdmlrein$. The galaxy sample as well as
  the colour coding is the same as in Fig. \ref{fig:GeneralPropt0}. }
\label{fig:SlopeDependence0}
\end{figure}

\begin{figure}
\centering
\includegraphics[width=8.5cm]{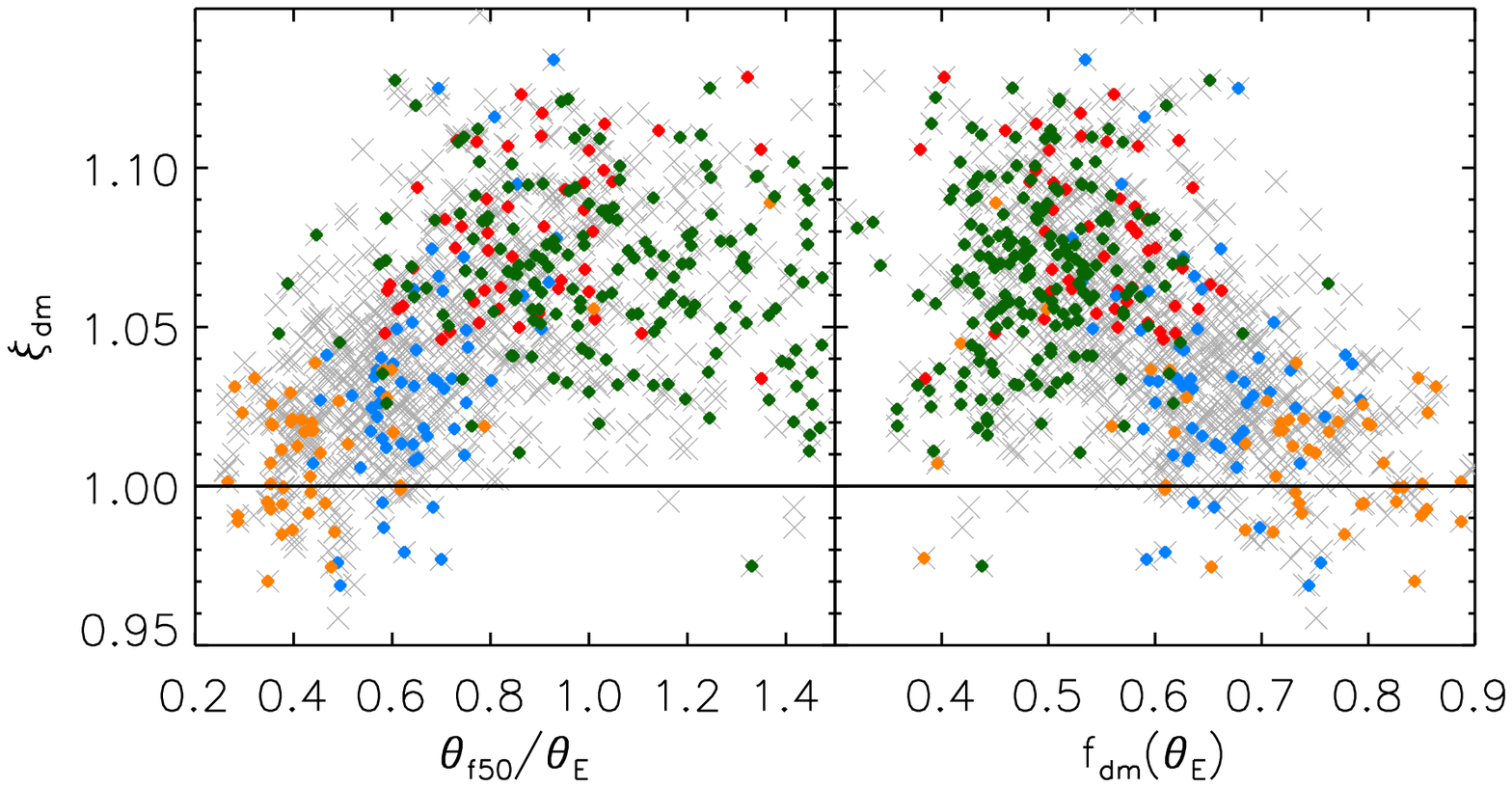}
\includegraphics[width=8.5cm]{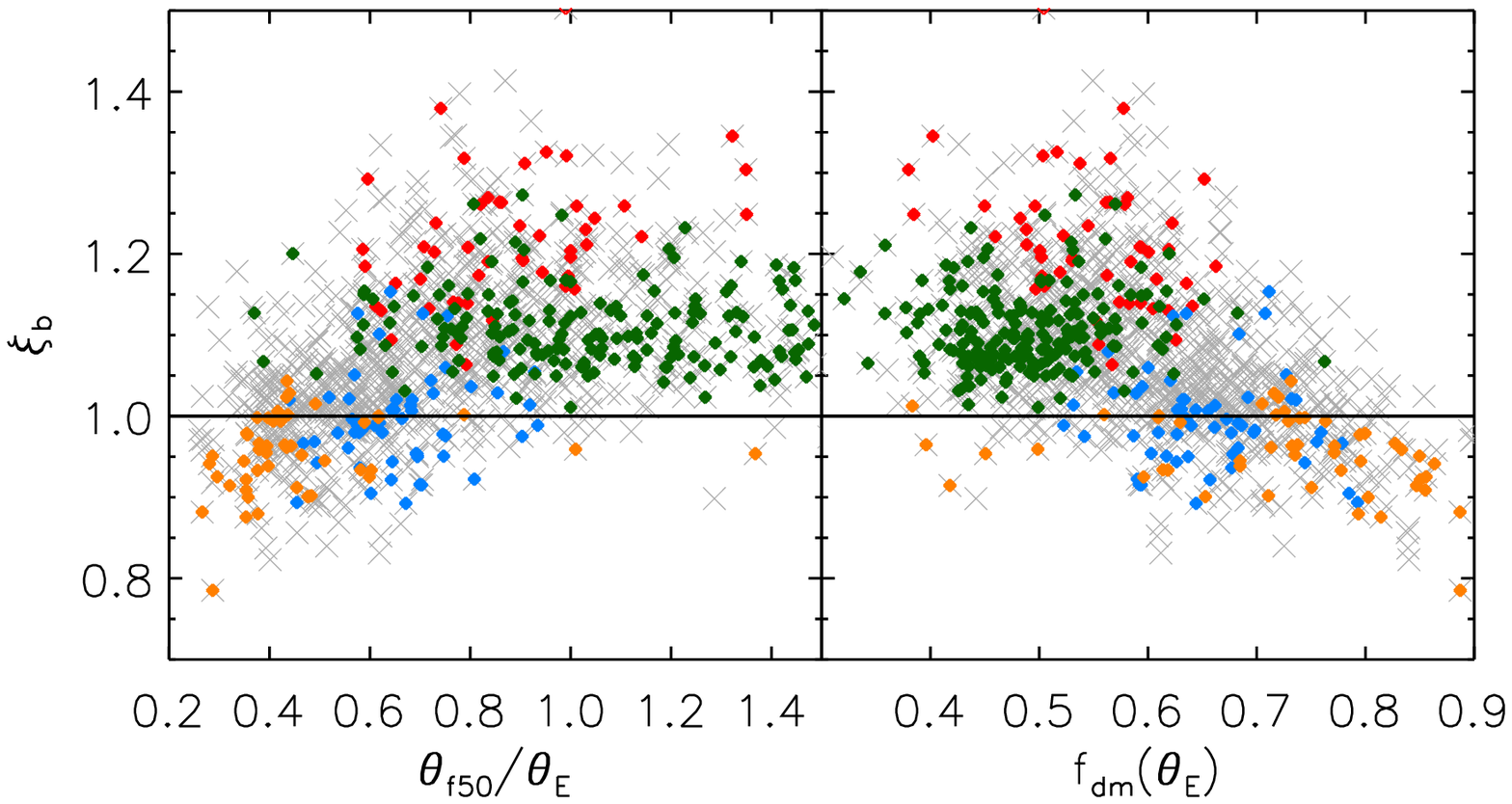}
\caption{Top left: dark matter curvature $\xi_{\rm dm}$ versus
  $\Rfdmlhf/\RE$. Top right: $\xi_{\rm dm}$ versus $\fdmlrein$.
  Bottom left: baryonic curvature $\xi_{\rm b}$ versus
  $\Rfdmlhf/\RE$. Bottom right: $\xi_{\rm b}$ versus $\fdmlrein$. The
  galaxy sample as well as the colour coding is the same as in
  Fig. \ref{fig:GeneralPropt0}. }
\label{fig:XiDependence0}
\end{figure}

\begin{figure}
\centering
\includegraphics[width=8.0cm]{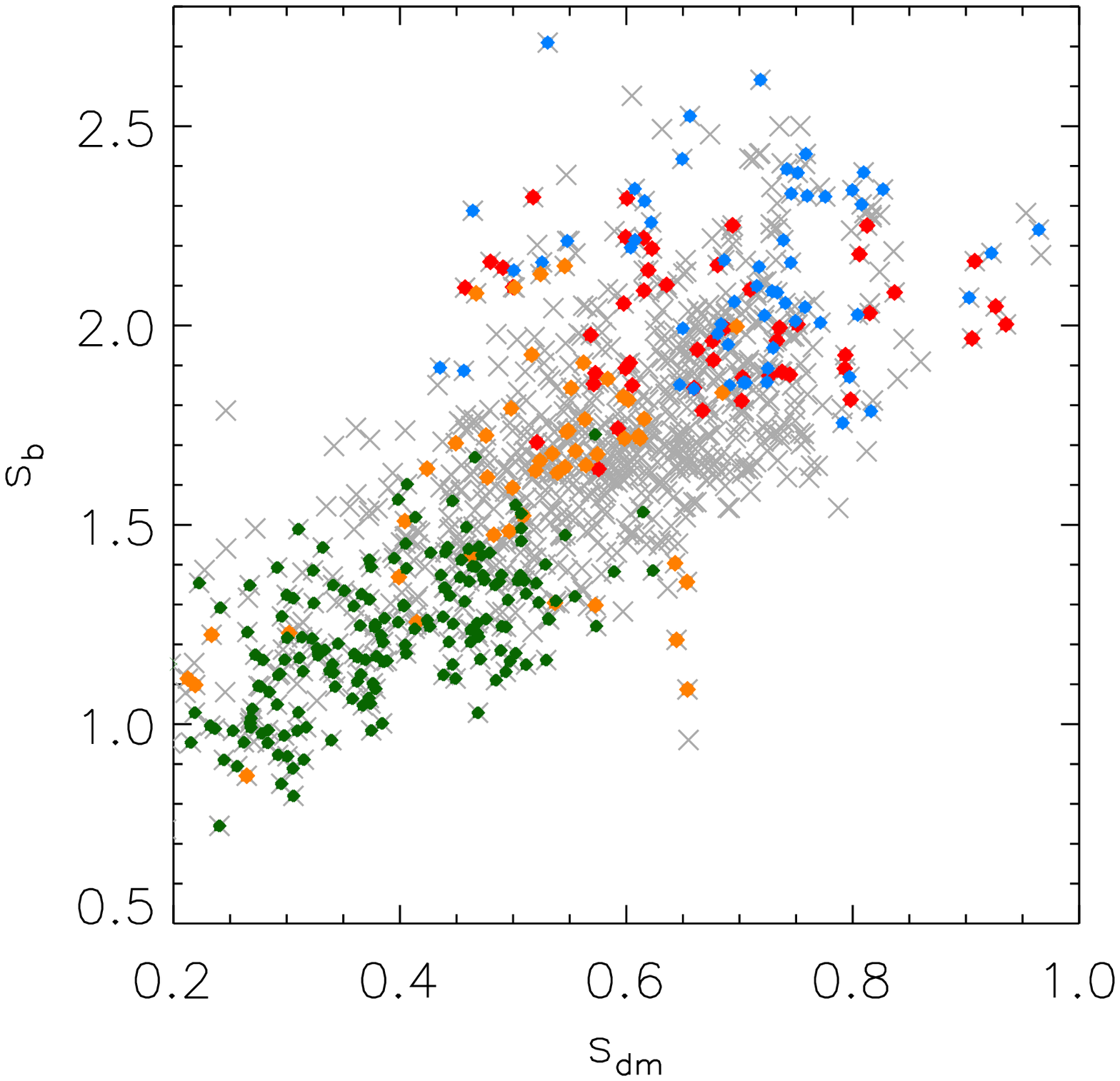}
\caption{The projected baryonic slope $s_{\rm b}$ versus dark matter
  slope $s_{\rm dm}$, measured between $\theta_1=0.5\,\RE$ and
  $\theta_2=1.5\,\RE$. The galaxy sample and the colour coding are the
  same as in Fig. \ref{fig:GeneralPropt0}. }
\label{fig:SDMvsSB}
\end{figure}

Fig. \ref{fig:SlopeDependence0} and \ref{fig:XiDependence0} show the
dependencies of the slopes and curvatures on $\Rfdmlhf/\RE$ and
$\fdmlrein$, which are two key quantities that are closely related to
different profiles of subgroups. In particular, $\Rfdmlhf/\RE$
describes where, with respect to $\RE$, the fraction of the projected
dark matter catches up and the distribution intercepts that of
baryons. Different values of $\Rfdmlhf/\RE$ reflect different parts of
the profile that strong lensing probes: a smaller $\Rfdmlhf/\RE$ means
that the strong lensing region ($\sim\RE$) is at larger radii than the
baryon-dark matter interception radius $\Rfdmlhf$, and thus the dark
matter fraction $\fdmlrein$ is larger. As indicated by the colour
coding, in the four highlighted subgroups, strong lensing probes
different parts of the density profile, from much further-out in
subgroup III (orange) to most closer-in in subgroup IV (green).

As can be seen from Fig. \ref{fig:SlopeDependence0}, the central
density profile of the dark matter distribution is always shallower
than isothermal, i.e., $s_{\rm dm}<1$; while that of the baryonic
matter in most cases is considerably steeper, i.e., $s_{\rm b}>1$. From
subgroup IV (green) to I (red), to II (blue) and to III (orange), as
strong lensing probes increasingly larger radii of the galaxy, a
significant profile steepening followed by a slope flattening develops
with increasing $\fdmlrein$ (or decreasing $\Rfdmlhf/\RE$); this is
seen for both dark matter and baryonic components in similar fashion.

In order to study the correlation between dark matter and baryonic 
slopes in the strong lensing region, Fig. \ref{fig:SDMvsSB} shows the
projected baryonic slope $s_{\rm b}$ as a function of dark matter
slope $s_{\rm dm}$ for the galaxy sample with $z_{\rm d}=0.2$ and
$z_{\rm s}=1.5$. In fact a strong correlation between the two slopes is
found for all of our samples with different $z_{\rm d}-z_{\rm s}$
combinations. This is a consequence of the strong interplay between
dark matter and baryons in the central regions of galaxies.

The dependencies of the curvature parameters, as presented in
Fig. \ref{fig:XiDependence0}, also exhibit similar distributions
between dark matter and baryons. Once again, as the strong lensing
region ``moves out'' (i.e., $\Rfdmlhf/\RE$ decreasing), a slight
increase (with concave curvature) followed by a decreasing trend (to
becoming convex) develops from subgroup IV (green) to I (red), to II
(blue) and to III (orange). 

Figures \ref{fig:SlopeDependence0} and \ref{fig:XiDependence0}
indicate the general trend of the (central and logarithmic) surface
density profiles of both dark matter and baryonic distributions: with
an increasing radius from the centre of a galaxy, a shallower and
concave-upward profile in the inner region (probed by strong lensing
of subgroup IV in green) is followed by a much steeper profile at
intermediate radii (now probed by strong lensing of subgroup I in red
and II in blue), and eventually is turning into a (slightly) shallower
profile again with convex curvature at larger radii (now probed by
strong lensing of subgroup III in green). Strong lensing therefore
probes different regions of such matter density distributions. In a
way, the distinctive profiles shown in Fig. \ref{fig:DensityProfile}
of galaxies in the four different subgroups are actually ``snapshots''
taken at different parts of the density distributions.


\subsection{Total density profiles}


\begin{figure}
\centering
\includegraphics[width=8.5cm]{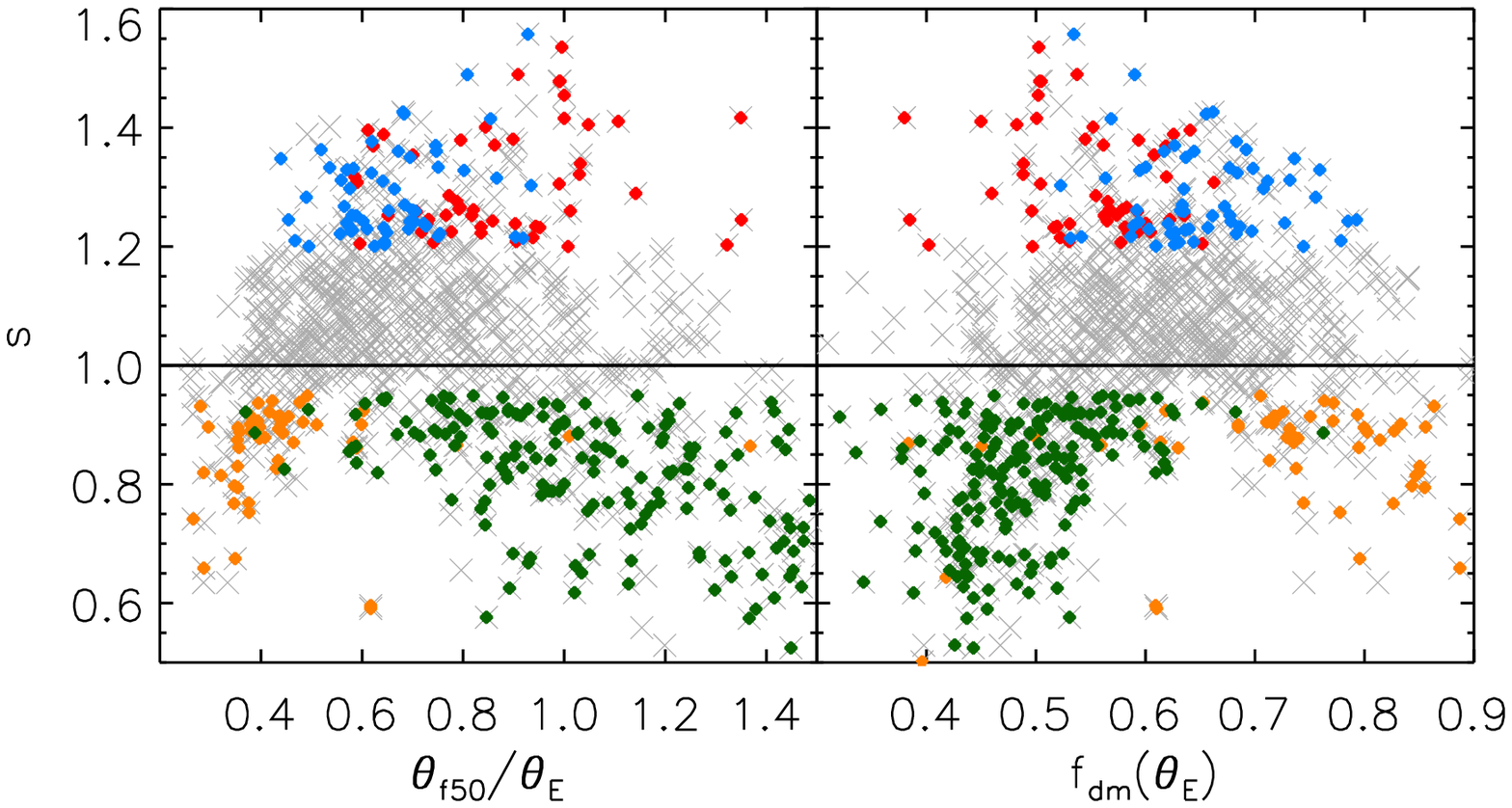}
\includegraphics[width=8.5cm]{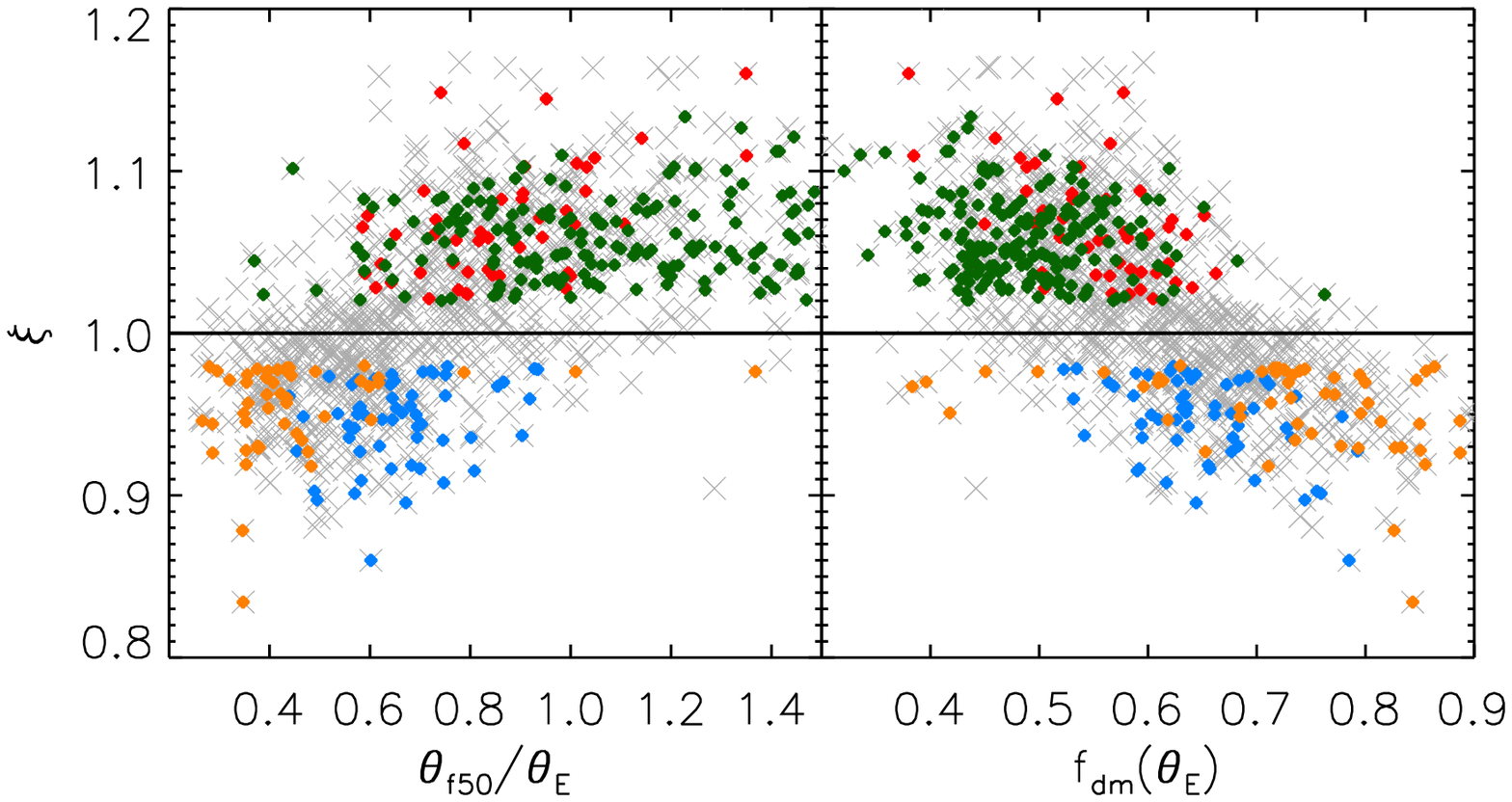}       
\caption{Top panels: $s$ versus $\Rfdmlhf/\RE$ (left) and $s$ versus
  $\fdmlrein$ (right). Bottom panels: $\xi$ versus $\Rfdmlhf/\RE$
  (left) and $\xi$ versus $\fdmlrein$ (right). The galaxy sample and
  the colour coding are the same as in Fig. \ref{fig:GeneralPropt0}.
}
\label{fig:TotProfDependence}
\end{figure}

We next present the results of the total (surface) density
profiles. Fig. \ref{fig:TotProfDependence} shows the distributions of
$s$ versus $\Rfdmlhf/\RE$ (top left), $s$ versus $\fdmlrein$ (top
right), $\xi$ versus $\Rfdmlhf/\RE$ (bottom left) and $\xi$ versus
$\fdmlrein$ (bottom right). Compared with
Fig. \ref{fig:SlopeDependence0} and \ref{fig:XiDependence0}, the
distributions of the total density profile parameters have maintained
the trends that are seen for both the dark matter and baryonic
components.

Strong gravitational lensing probes different parts of the total
density profile. This is the direct reason why we see a collection of
various profile behaviours. Among the four subgroups, the most massive
subgroup -- the one with shallower and convex density profiles (III,
in orange) -- have the smallest $\Rfdmlhf/\RE$. In particular their
$\Rfdmlhf\la0.5\RE$ indicates that their dark matter-baryon
interception radii are located within the inner boundaries of the
strong lensing regions. As a result, the total density profile
measured within the strong lensing regions is strongly dominated by
dark matter, whose distribution is considerably shallower than
isothermal (see Fig. \ref{fig:SlopeDependence0}).  In comparison,
subgroup II galaxies (in blue) have $\Rfdmlhf\ga0.5\RE$, i.e., the
dark matter-baryon interception radii now exceed the inner radii of
the strong lensing region, so that baryons can dominate the total
density distribution at $\theta_1$ (see Fig. 3), resulting in the
total slope $s$ steeper than isothermal. For both subgroups, a large
fraction of the measured density distribution is dominated by dark
matter; however, baryons contribute to a marked density upturn around
the inner boundary of the strong lensing region, resulting in a convex
curvature ($\xi<1$) of the total profile.

In comparison, subgroup I (in red) and IV (in green) have even lower
$\fdmlrein$ and larger $\Rfdmlhf/\RE$, which means that baryons
heavily dominate the strong lensing region in these cases. The
behaviour of the total density profile, therefore, largely depends on
how the baryonic matter is distributed. The strong lensing regions of
subgroup IV galaxies are much closer to (the projected) galaxy centres
than their subgroup I counterparts. This can be seen from larger
$\Rfdmlhf/\RE$ of the former than of the latter. As shown in
Sect. 4.1, both dark and baryonic matter distributions are much
shallower in the inner regions than at larger radii (see
Fig. \ref{fig:SlopeDependence0}). As a result, the total density
profiles of subgroup IV are shallower than of subgroup I galaxies;
while both have concave curvatures in the strong lensing regions
(i.e., $\xi_{\rm dm}>1$ and $\xi_{\rm b}>1$, see
Fig. \ref{fig:XiDependence0}).


\begin{figure*}
\centering
\includegraphics[width=16cm]{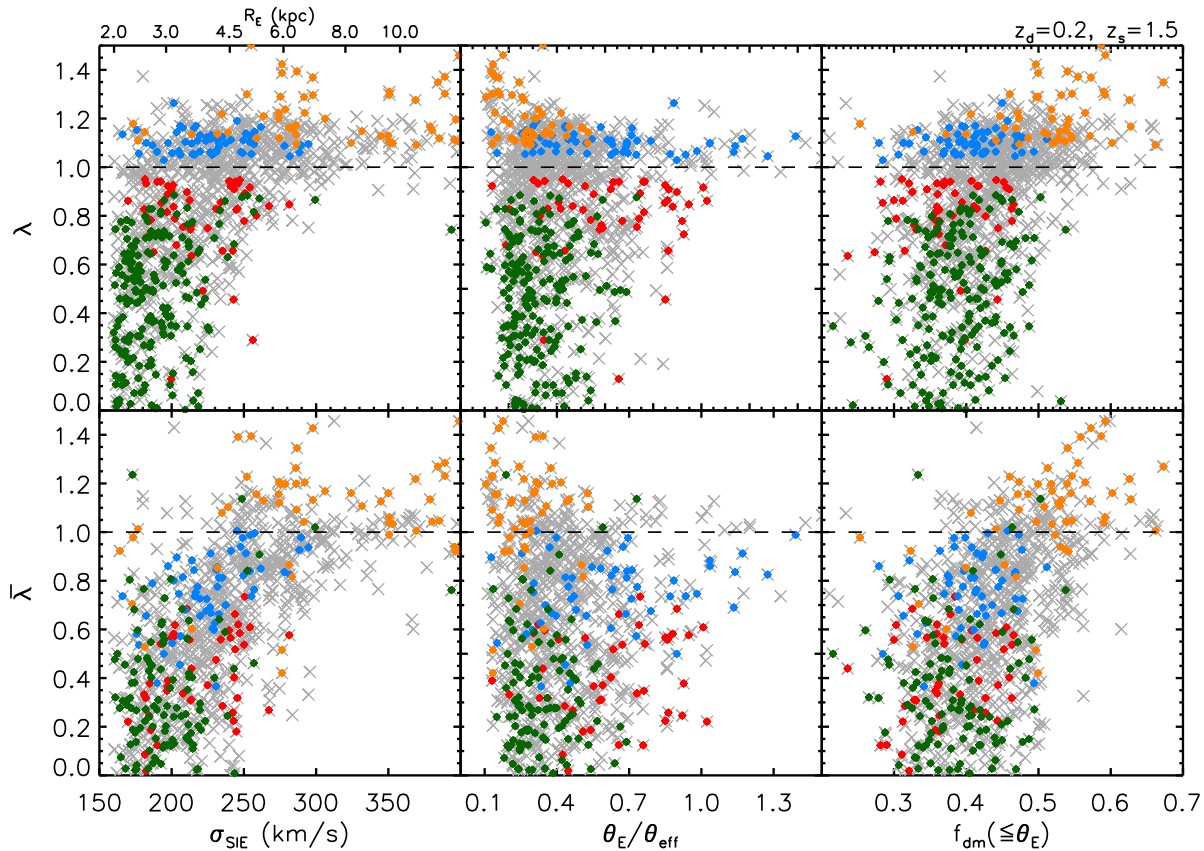}
\caption{The distributions of $\lambda$ (top) and $\bar\lambda$
  (bottom) versus $\sigma_{\rm SIE}$ (left), versus $\RE/\Reff$
  (middle) and versus $\fdmbrein$ (right). The sample as well as the
  colour coding is the same as in
  Fig. \ref{fig:GeneralPropt0}. Similar distributions are also seen
  for all other studied $z_{\rm d}-z_{\rm s}$ combinations. The upper
  axis of the top-left panel indicates the physical Einstein radius.}
\label{fig:LambdaDependence119}
\end{figure*}

\section{The distributions of $\lambda$ and the consequence for $H_0$ measurements}
\label{sec:lambda}

We have shown in Sect.~\ref{sec:Illustris} and \ref{sec:slope} that
galaxies can have a significant concavity/convexity compared to power
law density profiles. Because of the existence of the MST, the lens
modelling of these galaxies using a power-law density profile will
yield a multiplicative bias $\lambda$ on $H_0$. We study hereafter the
distributions of $\lambda$ resulting from the MST for the various
samples of Illustris galaxies. Following our notation, the MST applied
to the density profile $\kappa$ (via Eq.~\ref{eq:MST}) such that the
curvature parameter becomes unity, i.e., $\xi_\lambda=1$, yields a
bias $\lambda$ and a transformed profile $\kappa_{\lambda}$ with a
slope $s_{\lambda}$, while the MST on $\bar{\kappa}$ (via
Eq.~\ref{eq:MSTofKbar}) is characterized by $\bar{\lambda}$ and
$\bar{s}_{\lambda}$.

In Fig. \ref{fig:LambdaDependence119} we present three sets of
distributions, i.e., $\lambda$ and $\bar{\lambda}$ as a function of
$\sigma_{\rm SIE}$ (left), $\RE/\Reff$ (middle), and as a function of
$\fdmbrein$ (right), for the same galaxy sample as used before (where
$z_{\rm d}=0.2$ and $z_{\rm s}=1.5$). We see that the distributions of
$\lambda$ and $\bar{\lambda}$ span a wide range of values from 0 to
1.5 and exhibit a large scatter without a strong dependence on any of
the observables above. Figures displayed in Appendix~\ref{AppendixC}
show that similar distributions are also present for all other studied
$z_{\rm d}-z_{\rm s}$ combinations. It is however noticeable that for
galaxies with $200\kms < \sigma_{\rm{SIE}} < 300\kms$, the mean
multiplicative bias $\left<\lambda\right>$ is generally close to unity
with deviations not larger than $\sim20\%$ at lower
$\sigma_{\rm{SIE}}$, while for $\bar{\lambda}$ the mean deviation can
be as large as $\sim 50\%$. In all cases, an rms of $10\%-30\%$ is
present. This emphasizes that the power-law assumption in lens
modelling is highly questionable, as it will create non-negligible
biases in the derived values of $H_0$. However, we note that the
observed ``time delay lenses'' have generally large velocity
dispersion ($\sigma_{\rm SIE} \ga 250~\kms$), for which $\lambda$ is
close to unity on average, though with a $10\%-20\%$ scatter.

\begin{figure*}
\centering
\includegraphics[width=16cm]{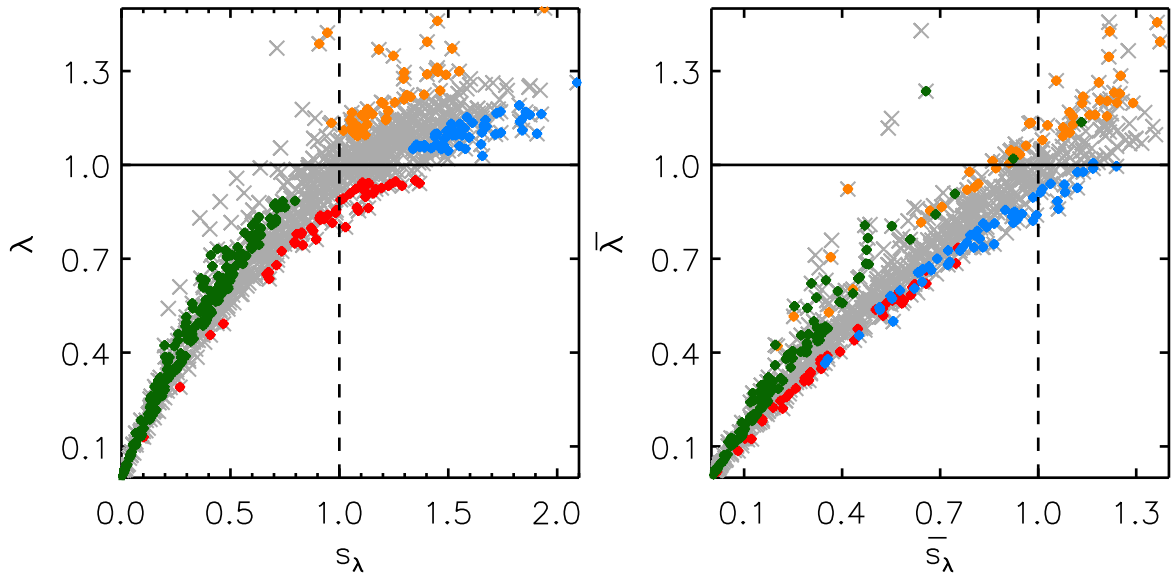}
\caption{The distributions of $\lambda$ versus $s_{\lambda}$ (left)
  and $\bar{\lambda}$ versus $\bar{s}_\lambda$ (right). The sample as
  well as the colour coding is the same as in
  Fig. \ref{fig:GeneralPropt0}. Similar distributions are also seen
  for all other studied $z_{\rm d}-z_{\rm s}$ combinations (given in
  the Appendix~\ref{AppendixC}). The solid lines indicate where
  $\lambda=1$ and where $\bar{\lambda}=1$; while the dashed lines
  indicate where $s_{\lambda}=1$ and where $\bar{s}_{\lambda}=1$. }
\label{fig:LambdavsSB4AT}
\end{figure*}

Interestingly, when correlating $\lambda$ with $s_{\lambda}$ (and
$\bar{\lambda}$ with $\bar{s}_{\lambda}$), which is the ``measured''
mean slope using power-law lens models, we see a much clearer
systematic trend with smaller scatters. This is shown in
Fig. \ref{fig:LambdavsSB4AT}, where the distributions of $\lambda$
versus $s_{\lambda}$ (left) and $\bar{\lambda}$ versus
$\bar{s}_{\lambda}$ (right) are presented. Such a tight correlation is
also seen for all other samples with different $z_{\rm d}-z_{\rm s}$
combinations (see Fig. \ref{fig:LambdaDistribution2}).

A striking result from Fig.~\ref{fig:LambdavsSB4AT} is that the
$s_{\lambda}$ (and $\bar{s}_\lambda$) distribution of galaxies with
$\lambda\approx1$ (and $\bar{\lambda}\approx1$) peaks at
$s_{\lambda}\approx1$ (and $\bar{s}_\lambda\approx1$); and the
$\lambda$ (and $\bar{\lambda}$) distribution of galaxies with
$s_{\lambda}\approx1$ (and $\bar{s}_{\lambda}\approx1$) peaks at
$\lambda\approx1$ (and $\bar{\lambda}\approx1$). This could be
potentially used to select time-delay samples which minimize the
impact of the MST on $H_0$ under the power-law assumption.

We have further investigated such a possibility. For each of our lens
samples, we have defined three subsamples that satisfy (1)
$s_{\lambda}\in[1-\Delta_s,~ 1+\Delta_s]$; (2)
$\bar{s}_{\lambda}\in[1-\Delta_s,~ 1+\Delta_s]$ and (3) both
$s_{\lambda}\in[1-\Delta_s,~ 1+\Delta_s]$ and
$\bar{s}_{\lambda}\in[1-\Delta_s,~ 1+\Delta_s]$, respectively.  The
slope span $\Delta_s$ is set to be 0.1, i.e., the ``measured'' mean
slopes (under the power-law assumption) are very close to
isothermal. Table \ref{tab:SelectedLambda} summarizes the statistical
properties of $\lambda$ and $\bar{\lambda}$ distributions for the
subsamples. In all the cases using these subsamples, both the systematic
deviation from unity ($\la5\%$) and the scatter ($\la10\%$) are much
smaller than those for the general samples.


\begin{table*}
\centering
\caption{A summary of the statistical properties of $\lambda$ and
  $\bar{\lambda}$ for galaxy subsamples that satisfy
  $s_{\lambda}\in[0.9,~1.1]$ and/or $\bar{s}_{\lambda}\in[0.9,~1.1]$:}
\begin{minipage} {\textwidth}
\begin{tabular}{c c c c | c c c} \hline\hline
Sample sets & \multicolumn{3}{c|}{$z_{\rm s}=1.5$} &
\multicolumn{3}{c}{$z_{\rm d}=0.6$} \\
\hline
Redshifts & ~~~$z_{\rm d}=0.2$~~~ & ~~~$z_{\rm d}=0.4$~~~ & ~~~$z_{\rm d}=0.8$~~~ &
~~~$z_{\rm s}=1.0$~~~ & ~~~$z_{\rm s}=1.5$~~~ & ~~~$z_{\rm s}=3.0$~~~ \\\hline
\multicolumn{7}{c}{Subsample I: $s_{\lambda}\in[0.9,~1.1]$} \\\hline
Number of galaxy projections & 142 & 210 & 174 & 146 & 223 & 296 \\
Mean $\lambda$         & 1.00 & 0.98 & 0.97 & 0.97 & 0.97 & 0.97 \\  
Median $\lambda$       & 0.99 & 0.97 & 0.96 & 0.95 & 0.96 & 0.97 \\
Standard deviation $\sigma_\lambda$ & 0.11 & 0.09 & 0.10 & 0.12 & 0.10 & 0.08 \\\hline
\multicolumn{7}{c}{Subsample II: $\bar{s}_{\lambda}\in[0.9,~1.1]$} \\\hline
Number of galaxy projections & 110 & 167 & 148 & 102 & 184 & 278 \\
Mean $\bar{\lambda}$          & 0.98 & 0.98 & 0.97 & 0.96 & 0.97 & 0.98 \\
Median $\bar{\lambda}$        & 0.98 & 0.98 & 0.95 & 0.96 & 0.97 & 0.97 \\
Standard deviation $\sigma_{\bar{\lambda}}$ & 0.08 & 0.09 & 0.11 & 0.09 & 0.10 & 0.14 \\\hline
\multicolumn{7}{c}{Subsample III: $s_{\lambda}\in[0.9,~1.1]$ and $\bar{s}_{\lambda}\in[0.9,~1.1]$} \\\hline
Number of galaxy projections & 19 & 36 & 21 & 19 & 37 & 63 \\
Mean $\lambda$                & 1.04 & 1.02 & 1.04 & 1.06 & 1.01 & 1.01 \\
Median $\lambda$              & 1.03 & 1.02 & 1.01 & 1.04 & 1.02 & 1.01 \\
Standard deviation $\sigma_\lambda$        & 0.06 & 0.06 & 0.10 & 0.08 & 0.05 & 0.07 \\
Mean $\bar{\lambda}$          & 1.02 & 1.02 & 1.04 & 1.06 & 1.01 & 1.02 \\
Median $\bar{\lambda}$        & 1.01 & 1.01 & 1.02 & 1.04 & 1.01 & 1.01 \\
Standard deviation $\sigma_{\bar{\lambda}}$ & 0.05 & 0.06 & 0.12 & 0.08 & 0.07 & 0.08 \\\hline
\end{tabular}
\end{minipage}
\label{tab:SelectedLambda}
\end{table*}


The statistical results presented above indicate that the power-law
assumption will introduce non-negligible biases in $H_0$ measurements,
even though adopting power-law lens models is a common practice.
However, through a careful sample selection, e.g., selecting galaxies
that have larger velocity dispersions (i.e., large angular separations
of the multiple images) or those which have the ``measured'' profiles
close to isothermal under the power-law assumption, the use of simple
power-law models may yield estimates of $H_0$ which are biased only at
the few percent level -- compared to a much larger bias from the total
sample studied here. This work also shed light on such a possibility
in this direction.

\section{Discussion and conclusions}

The central regions of massive galaxies (i.e. typically the inner 5-10
kpc) contain dark and baryonic matter in roughly similar
amounts. Although both components follow different density
distributions, it has been found that the total density profile in
that region is well approximated by a single power law (e.g., Koopmans
et al. 2006, Gavazzi et al, 2007). This has motivated the use of the
power-law model $\rho \propto r^{-\gamma'}$ as a generic model for
numerous galaxy-scale strong lensing studies. However, as emphasized
by Schneider \& Sluse (2013), a mass distribution $\kappa(\theta)$ can
be transformed into an approximate power law by means of an MST,
i.e. $\kappa_{\lambda}(\theta) =\lambda \kappa(\theta)+(1-\lambda)$,
even if the true mass profile shows considerable curvature.  Under a
MST, none of the lensing observables are modified except for the
product of the Hubble constant and the time delay $H_0\,\Delta t$,
which is scaled by the same factor $\lambda$ as the one characterizing
the MST. By assuming a power-law profile, one ``artificially'' breaks
the MSD and derives a biased estimate of $H_0$. In addition, the use
of a single quantity, i.e., the logarithmic slope $\gamma'$, to
characterize the density profile of galaxies may be misleading, since
the intrinsic density does not follow an exact power law. Therefore,
such a slope depends on the radius, and its comparison among galaxies
depends on the range of galactocentric radii over which it is measured
[note that Dutton \& Treu 2014 suggests the use of a mass-weighted
slope to alleviate the problem].

In this paper, we have used a statistical sample of mock lensing
galaxies from the first high-resolution cosmological-scale
hydrodynamic simulation -- the Illustris Project
(\citealt{Illustris2014Nat}, see also \citealt{Illustris2014MN,
  Genel2014Illustris, Nelson2015IllustrisDataRelease}) -- to test the
validity of the power-law assumption for the (central) surface density
distribution of strong gravitational lenses.  The simulation
reproduces extremely well numerous observational properties of real
galaxies. In particular, we find that the cumulative dark matter
fraction $\fdmbrein$ of the simulated lensing galaxies ranges from
30\% to 70\%, in good agreement with observations (Treu \& Koopmans
2004; Koopmans et al. 2006). The average isothermality of the profiles
derived for the simulated galaxies is also in agreement with various
observational studies (e.g., Rusin et al. 2003; Auger et al. 2010;
Sonnenfeld et al. 2013). A detailed analysis of the 3d density profile
in the inner regions of Illustris lenses will be presented in a
forthcoming paper (Xu et al., in preparation).

From the mock Illustris lensing galaxies at each of the studied lens
redshifts below $z=1$, we have taken a sample of about 400 plausible
lens galaxies, which have central velocity dispersion $\sigma
\geqslant 160\kms$ and are ``observed'' through their three
independent projections, yielding a sample of more than 1000 projected
density profiles. The slopes and curvatures of the projected density
profiles in regions probed by strong lensing (typically between 0.5
and 1.5 angular Einstein radii $\theta_E$) have been measured. We have
transformed these profiles into (local) power laws via an MST and
derived the distribution of the multiplicative bias $\lambda$ on $H_0$
implied by a power-law assumption of the density profile.

Our main findings are as follows:
\begin {itemize}


\item{The total surface density profile in the projected {\it central}
  regions of galaxies depends on radius, showing deviations from a
  power law following some generic trends. At smaller radii the
  profiles are shallow and concave upward. With increasing radius, the
  profiles gradually steepen, and eventually at larger radii they
  become marginally shallower and attain convex curvatures. The
  projected density distributions of both dark and baryonic matter
  follow such a general trend, while the latter is more concentrated
  (and steeper) than the former. Good correlations exist between the
  dark matter and baryonic density slopes throughout the projected
  radii and redshifts studied in this work, indicating the strong
  interplay between dark matter and baryons in central regions of
  galaxies.}

\item {In the region probed by strong lensing, the (projected) total
  density profile deviates significantly (by up to 15\%) from an exact
  power law, with a variety of slopes and curvatures. The main reason
  for the diverse measurements of the profile parameters is that
  strong lensing probes different parts of this density distribution;
  the measured slopes and curvatures depend on the relative
  contribution of dark and baryonic matter in regions where the
  measurements are made. In general dark matter dominates the strong
  lensing regions of massive galaxies, which therefore tend to have
  mean profiles (in the strong lensing region) shallower than
  isothermal. In contrast, lower-mass galaxies have non-negligible
  baryonic components within their strong lensing regions; the shapes
  of the mean total density profiles are thus largely determined by
  the baryonic distributions. }

\item{The curvature of the (logarithmic) density profile is
  considerable and produces a non-negligible bias in the measured $H_0$, if
  only lensing information is used. The statistical distribution of
  $\lambda$ (and $\bar{\lambda}$) derived from the Illustris strong
  lens sample spans from 0 to 1.5 and exhibits large scatter without
  a strong dependence on observationally-constrained galaxy
  properties, such as $\sigma_{\rm SIE}$, $\RE/\Reff$ and
  $\fdmbrein$. In particular, the mean deviation of $\lambda$ (and
  $\bar{\lambda}$) from unity can be as large as $20\%-50\%$ with a
  scatter of $10\%-30\%$ (rms) for galaxies with $\sigma_{\rm
    SIE}=200\kms-300\kms$. We note, however, that galaxies with derived
  slopes (under the power-law assumption) close to isothermal exhibit
  much smaller systematic deviation of $\la5\%$ from unity with a
  scatter of $\la10\%$ (rms). This could potentially be used to form a
  reliable sample of lensing galaxies for $H_0$ measurements in the
  upcoming big data era. }
\end {itemize}

We would like to stress that by finding an MST which approximates
$\kappa_\lambda$ as closely as possible by a power law, we in essence
approximate $\kappa$ locally by a power law plus a uniform mass
sheet. However, this is purely a mathematical description; this
uniform mass sheet is not assigned any physical meaning. In
particular, this mass sheet cannot (or only partly) be interpreted as
a convergence due to foreground and background material along the line
of sight. The latter contribution to the convergence can be
estimated/constrained using cosmological simulations and/or through
detailed studies of the lens environment or weak lensing (e.g.,
\citealt{Sherry2010B1608, Sherry2013J1131}).
Furthermore, even if a lens is perfectly fitted with $\kappa$ being a
power law plus uniform mass sheet, then the same is true for all
$\kappa_\lambda$ through MST. Hence, $\lambda$ remains undetermined.

It is also worth noting that one may try out different ``realistic''
model profiles for the mass distribution of lenses, such as the
composite ones used in \citet{SS13} or in \citet{Suyu2014JX1131}, in
order to assess the systematics due to the MST. However the principal
problem remains: what is the possible range of ``realistic'' mass
models? In other words, if a (possible composite) model $\kappa$ fits
the data, how can one rule out that the corresponding model
$\kappa_\lambda$ (with, say, $0.85\leqslant \lambda\leqslant 1.15$) is
not ``realistic''? The various density profiles shown in Fig.3 indicate
that real galaxies most likely exhibit a fairly large range of density
profiles, which makes the classification of models profiles into
``realistic'' and less realistic ones fairly difficult, if not all
impossible.

Thereby biased estimates of $H_0$ are not fundamentally due to the
power-law assumption of the lens model but are caused by the MSD
itself, whose impact cannot be eliminated simply by studying the
properties of material in the surrounding light cone, or through the
adoption of more sophisticated lens models even when extra constraints
are available from extended image configurations. In order to obtain a
reliable estimate of $H_0$ using gravitational lensing, one must first
break the MSD.

One possibility of doing so is to use the fact that MST also modifies
image magnification $\mu\rightarrow \mu/\lambda^2$. If the source
luminosity, and thus the absolute magnification, is known $\lambda$
can be fixed. For AGN as sources, as shown by
\citet{Bauer2012VariabilityLuminosity}, the variability-luminosity
relation can be used to estimate source luminosities but only in a
statistical way with large scatter. Alternatively, if Type Ia
supernovae are gravitationally lensed (\citealt{Chornock2013PS1,
  Quimby2013PS1, Quimby2014TypeIaLens}), direct measurements of $\mu$
would be possible due to their nature as approximate standard
candles. In such cases, the MSD can also be broken and the
multiply-lensed Type Ia supernovae become ideal laboratories to study
the true lens mass distribution as well as cosmological parameters
(\citealt{Kolatt1998IALens, Oguri2003IaLensBMSD, Bolton2003SNLens}).

A second way to break the MSD is to include independent information
about the lens mass distribution, e.g., from stellar dynamics (e.g.,
\citealt{BarnabeKoopmans2007, Barnabe2009II, Barnabe2011III,
  TK2002PG1115, Sherry2010B1608, Sherry2013J1131, Suyu2014JX1131}).
The problem is that the current accuracy (typically 10\%) in the
velocity dispersion measurements may not yet be sufficient to break
the MSD, not to mention the uncertainty due to the anisotropy of
stellar orbits, which can lead to a systematic error on the slope at a
level of $\sim5\%$ (\citealt{Agnello2013}). Further investigations in
this regard were not included in this analysis but are greatly
encouraged.

\section*{ACKNOWLEDGEMENTS}
The authors would like to thank Stefan Hilbert, Sherry Suyu, Malte
Tewes, James Nightingale and an anonymous referee for very useful
comments. DDX thanks the HITS fellowship. DS acknowledges support from
a {\it {Back to Belgium}} grant from the Belgian Federal Science
Policy (BELSPO), and partial funding from the Deutsche
Forschungsgemeinschaft, reference SL172/1-1. PS and VS acknowledge
support by the DFG through Transregio 33, ``The Dark Universe''. VS
also acknowledges support by the European Research Council under
ERC-StG EXAGAL-308037. LH acknowledges support from NASA grant
NNX12AC67G and NSF grant AST-1312095.

\onecolumn
\appendix

\section{Varying the range of the strong lensing region }
\label{AppendixA}

\begin{figure*}
\centering
\includegraphics[width=16cm]{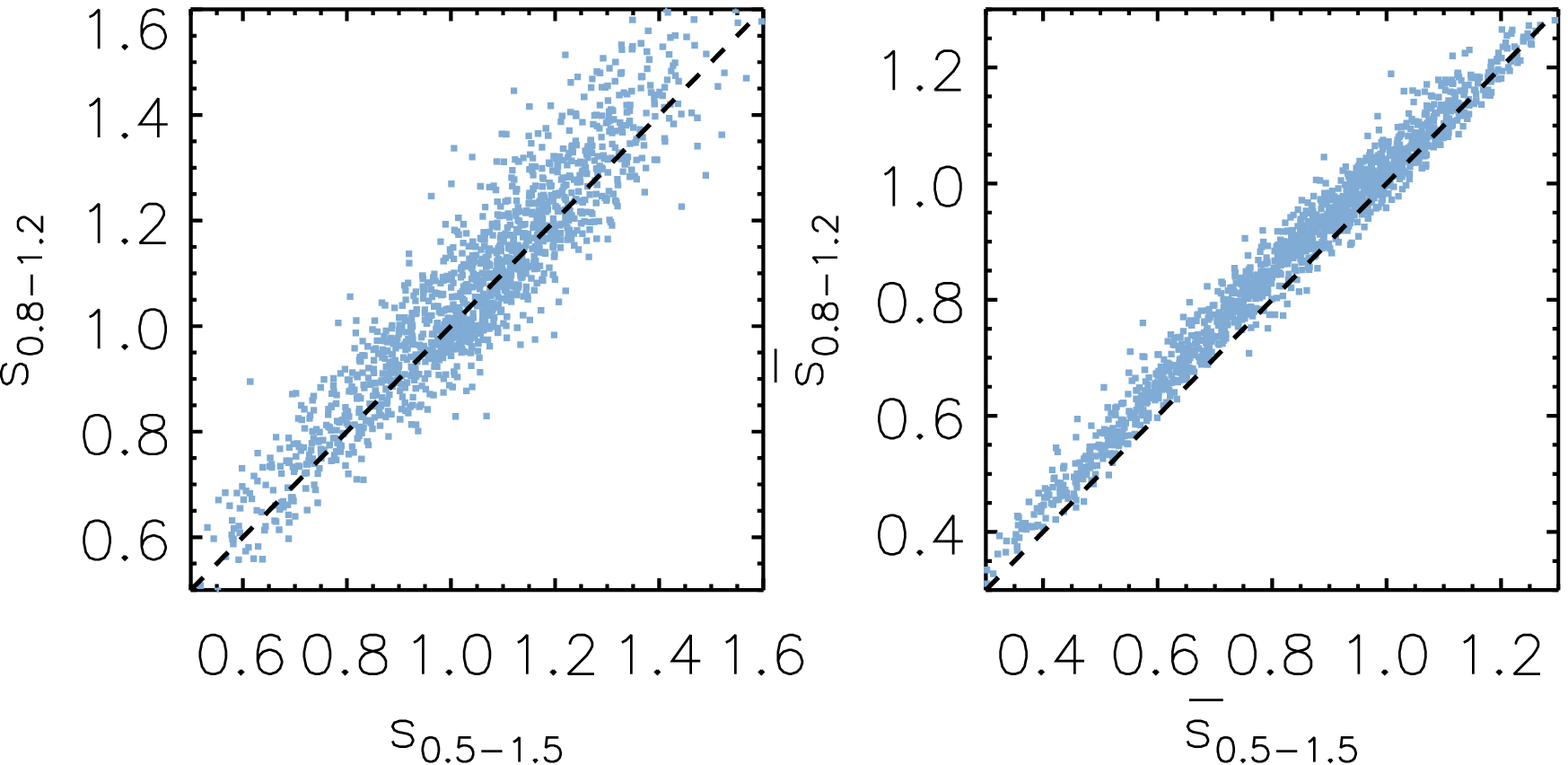}
\caption{The slopes defined within different radial ranges are
  compared for the lens sample with $z_{\rm d}=0.6$ and $z_{\rm
    s}=1.5$. On the left, the slopes of the local convergence
  $\kappa(\theta)$ measured between $\theta_1=0.8\RE$ and
  $\theta_2=1.2\RE$ are plotted versus those measured between
  $\theta_1=0.5\RE$ and $\theta_2=1.5\RE$; on the right the relation
  of the cumulative distribution $\bar{\kappa}(\leqslant\theta)$ is
  given.}
\label{fig:SlopesT12vary}
\end{figure*}

\begin{figure*}
\centering
\includegraphics[width=16cm]{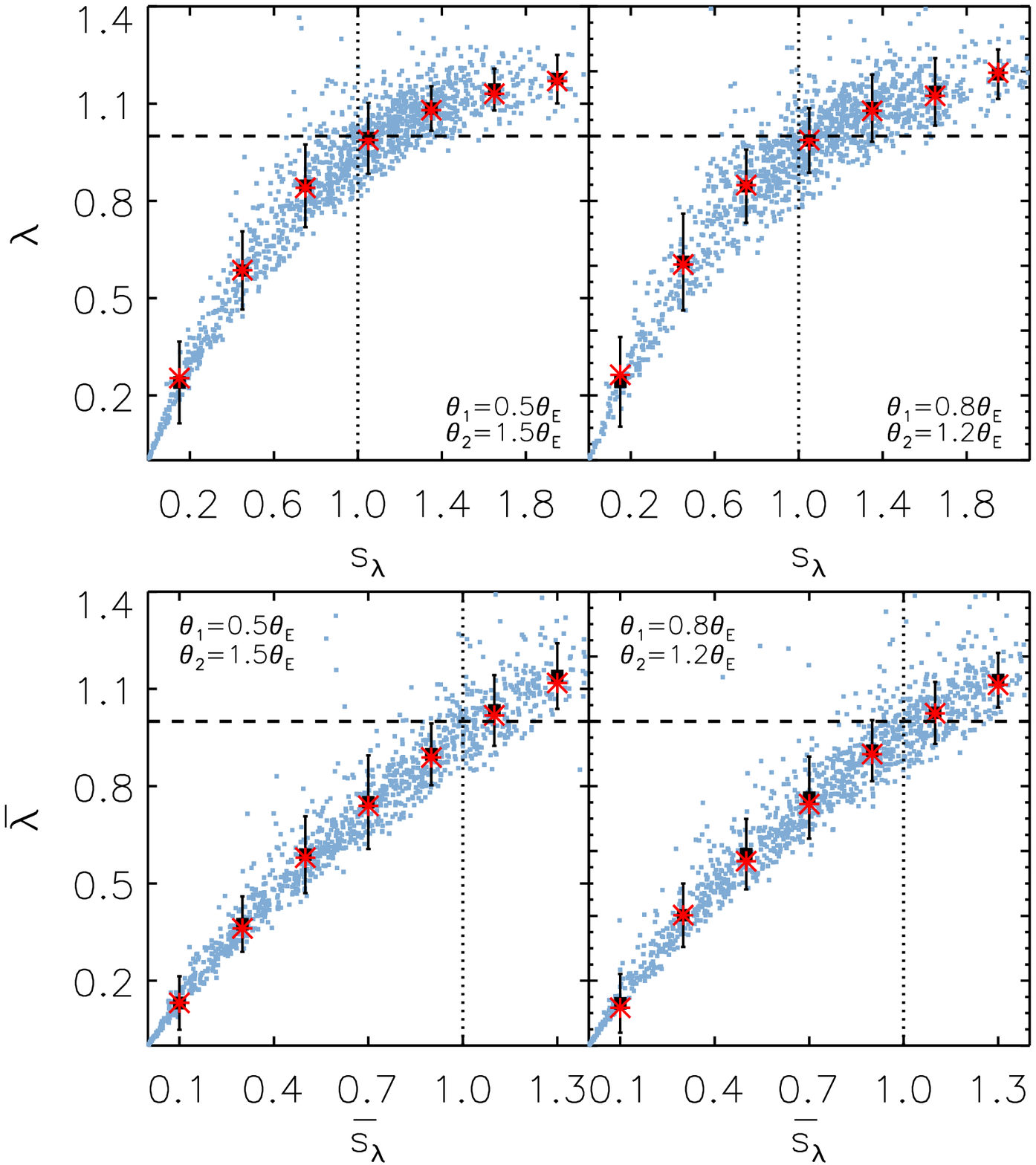}
\caption{The distributions of $\lambda$ versus $s_{\lambda}$ (top
  panels) and $\bar{\lambda}$ versus $\bar{s}_{\lambda}$ (bottom
  panels) for the lens sample with $z_{\rm d}=0.6$ and $z_{\rm
    s}=1.5$: the left-hand side panel shows the distributions
  calculated assuming $\theta_1=0.5\RE$ and $\theta_2=1.5\RE$; for
  comparison, the results for which $\theta_1=0.8\RE$ and
  $\theta_2=1.2\RE$ are adopted are given on the right-hand side. On
  top of the scattered data (blue dots), the black squares (and red
  stars) with the error bars indicate the mean (and median) and its
  standard deviation within each bin. The dashed lines indicate where
  $\lambda=1$ and where $\bar{\lambda}=1$; while the dotted lines
  indicate where $s_{\lambda}=1$ and where $\bar{s}_{\lambda}=1$.}
\label{fig:LambdaT12vary}
\end{figure*}

Different angular ranges $[\theta_1,~\theta_2]$ will not result in
exactly the same transformed power-law profiles and the associated
$\lambda$s. It is therefore interesting to consider if our final
statistical results strongly depend on the choice of the image
range. In Fig. \ref{fig:SlopesT12vary}, we present the slopes measured
between $\theta_1=0.5\RE$ and $\theta_2=1.5\RE$ and between
$\theta_1=0.8\RE$ and $\theta_2=1.2\RE$. The corresponding $\lambda$
distributions are shown in Fig. \ref{fig:LambdaT12vary}. As can be
seen, although measurements in these two cases are not exactly the
same on a one-to-one basis, the statistical distributions of $\lambda$
are independent of the image range sets. We have, therefore, set
$\theta_1=0.5\RE$ and $\theta_2=1.5\RE$ to calculate the relevant
quantities throughout the paper.

\section{Density profiles}
\label{AppendixB}
\begin{figure}
\centering
\includegraphics[width=8cm]{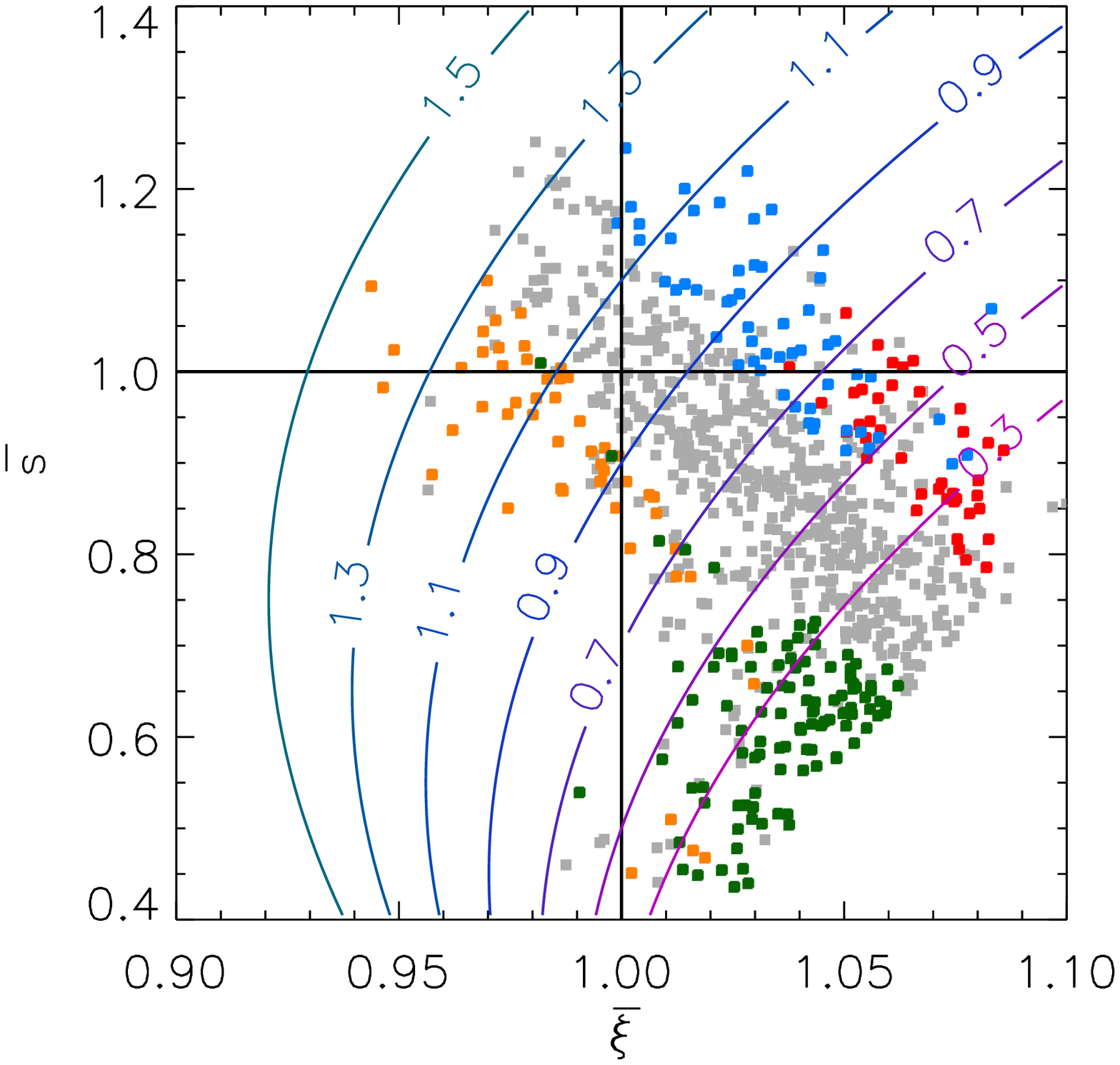}
\caption{The $\bar{s}-\bar{\xi}$ distribution of the selected lensing
  galaxies at $z_{\rm d}=0.2$, assuming $z_{\rm s}=1.5$. Four colours
  represent four subgroups divided according to $s$ and $\xi$ of the
  local convergence distribution $\kappa(\theta)$: subgroup I (red)
  are defined as galaxies with $s>1.2$, $\xi>1.02$; subgroup II (blue)
  are those with $s>1.2$, $\xi<0.98$; subgroup III (orange) represents
  galaxies with $s<0.95$, $\xi<0.98$; and subgroup IV (green) are
  those with $s<0.95$, $\xi>1.02$. Contours indicate where the
  transformed slopes $\bar{s}_{\lambda} =
  [0.3,~0.5,~0.7,~0.9,~1.1,~1.3,~1.5,~1.7]$. }
\label{fig:GeneralPropt0bar}
\end{figure}

\begin{figure*}
\centering
\includegraphics[width=16cm]{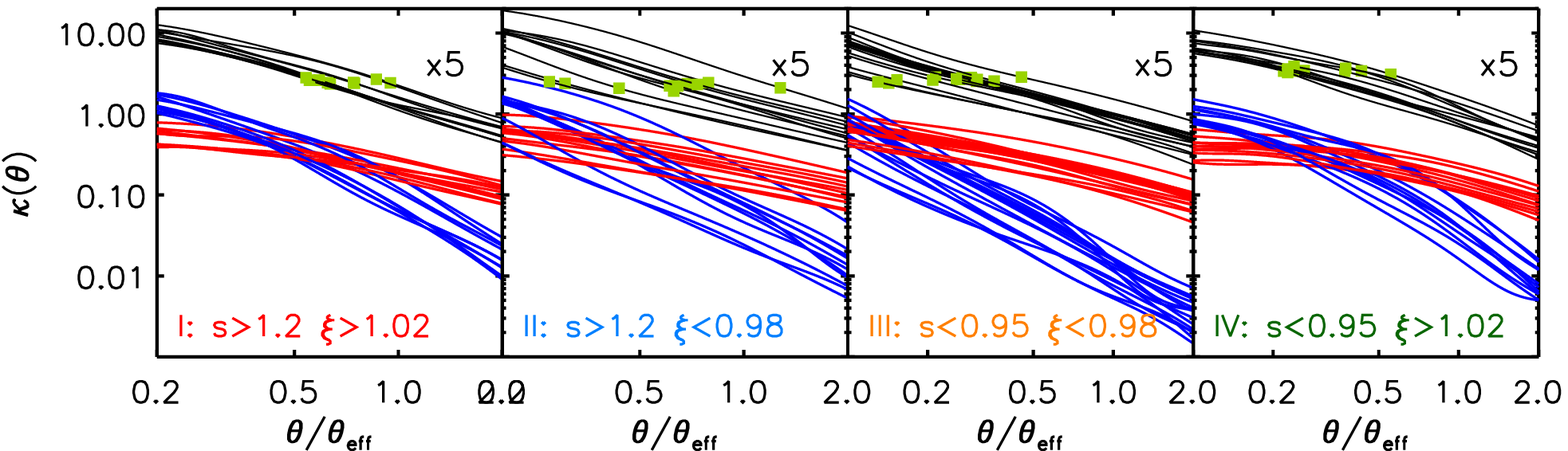}
\caption{The surface density profiles of galaxies from each subgroup
  of the lens sample presented in Fig. 3 are shown as normalized by
  galaxy effective radii $\theta_{\rm eff}$. The black curves show the
  total surface density distribution $\kappa(\theta)$, scaled up by a
  factor of 5 for clarity; the red and blue curves represent profiles
  of the projected dark matter $\kappa(\theta) f_{\rm dm}(\theta)$ and
  projected baryonic mass $\kappa(\theta) (1-f_{\rm dm}(\theta))$,
  respectively; green dots indicate where the Einstein radii are with
  respect to the effective radii.}
\label{fig:DensityProfile2Reff}
\end{figure*}

In this appendix, we show the density profiles of our galaxy
samples. Fig.~\ref{fig:GeneralPropt0bar} presents the shape parameter
distribution in the $\bar{s}-\bar{\xi}$ plane (see Sect.\ 2 for
definition). Fig.~\ref{fig:DensityProfile2Reff} shows the surface
density profiles of galaxies from each subgroup of the lens sample
presented in Fig. 3, which are now normalized by galaxy effective
radii $\theta_{\rm eff}$.

\section{The $\lambda$ distributions of all lens samples}
\label{AppendixC}
In this appendix, we present the distributions of $\lambda$ (and
$\bar{\lambda}$) as a function of $\sigma_{\rm SIE}$ and of the
``measured slope'' $s_{\lambda}$ (and $\bar{s}_{\lambda}$) for our
lens samples with different sets of $z_{\rm d}$ and $z_{\rm s}$
combinations.
 
\begin{figure*}
\centering
\includegraphics[width=12cm]{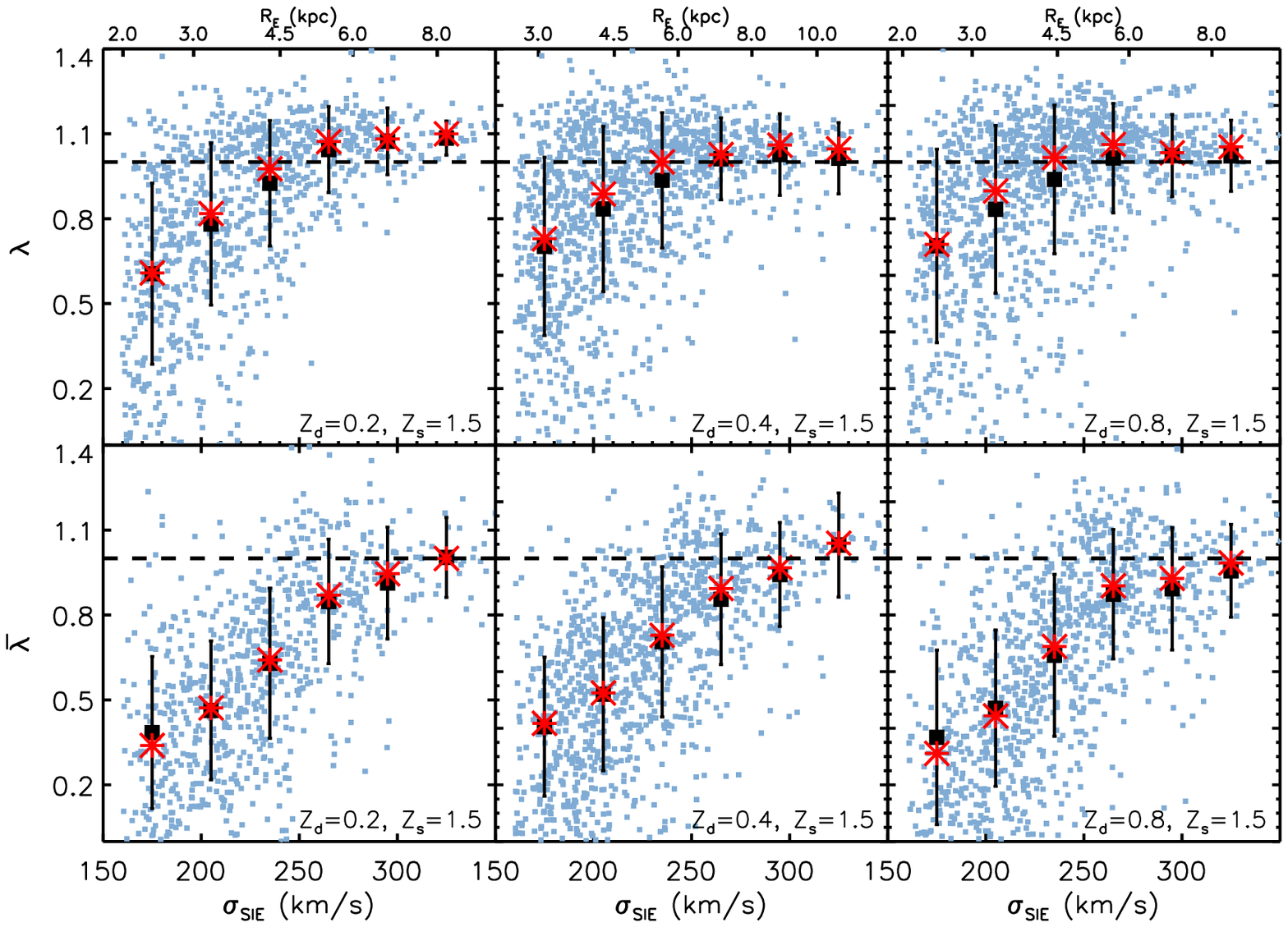}
\includegraphics[width=12cm]{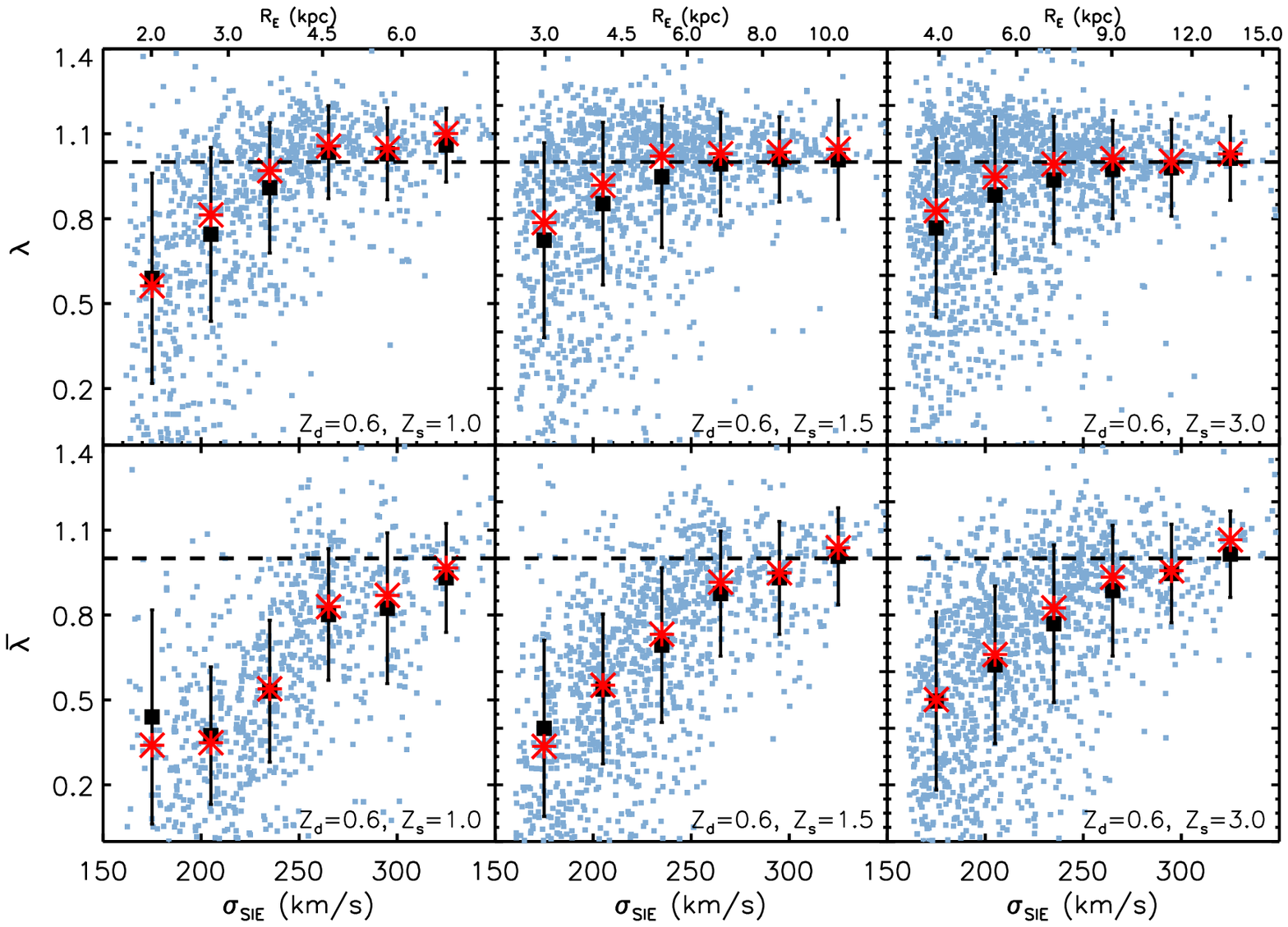}
\caption{Distributions of $\lambda$ (and $\bar{\lambda}$) versus
  $\sigma_{\rm SIE}$ for all six samples with $z_{\rm d}$ and $z_{\rm
    s}$ indicated in the panels. On top of the scattered data (blue
  dots), the black squares (and red stars) with the error bars
  indicate the mean (and median) and its standard deviation within
  each bin. The dashed lines indicate where $\lambda=1$ and where
  $\bar{\lambda}=1$; while the dotted lines indicate where
  $s_{\lambda}=1$ and where $\bar{s}_{\lambda}=1$.}
\label{fig:LambdaDistribution1}
\end{figure*}

\begin{figure*}
\centering
\includegraphics[width=12cm]{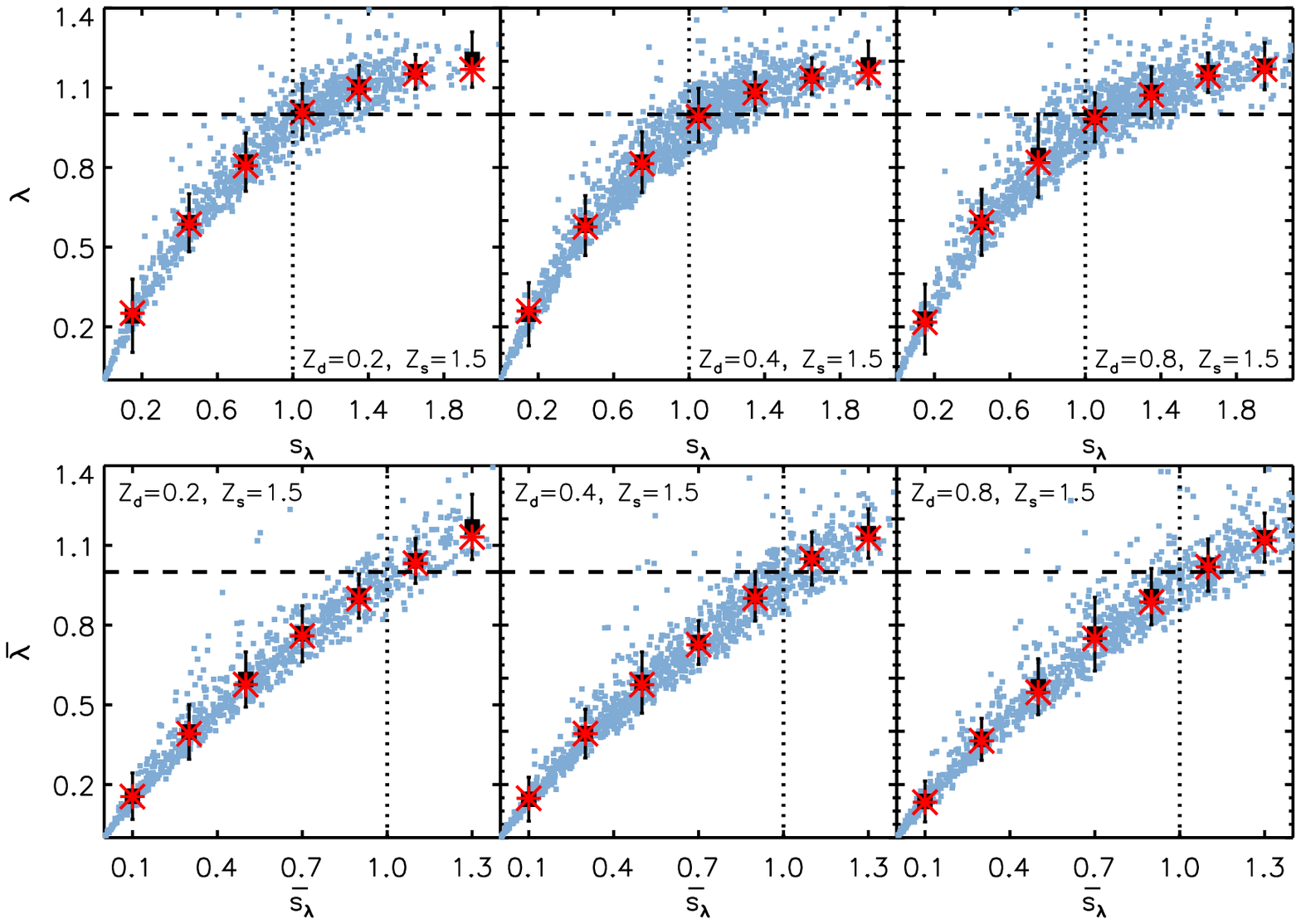}
\includegraphics[width=12cm]{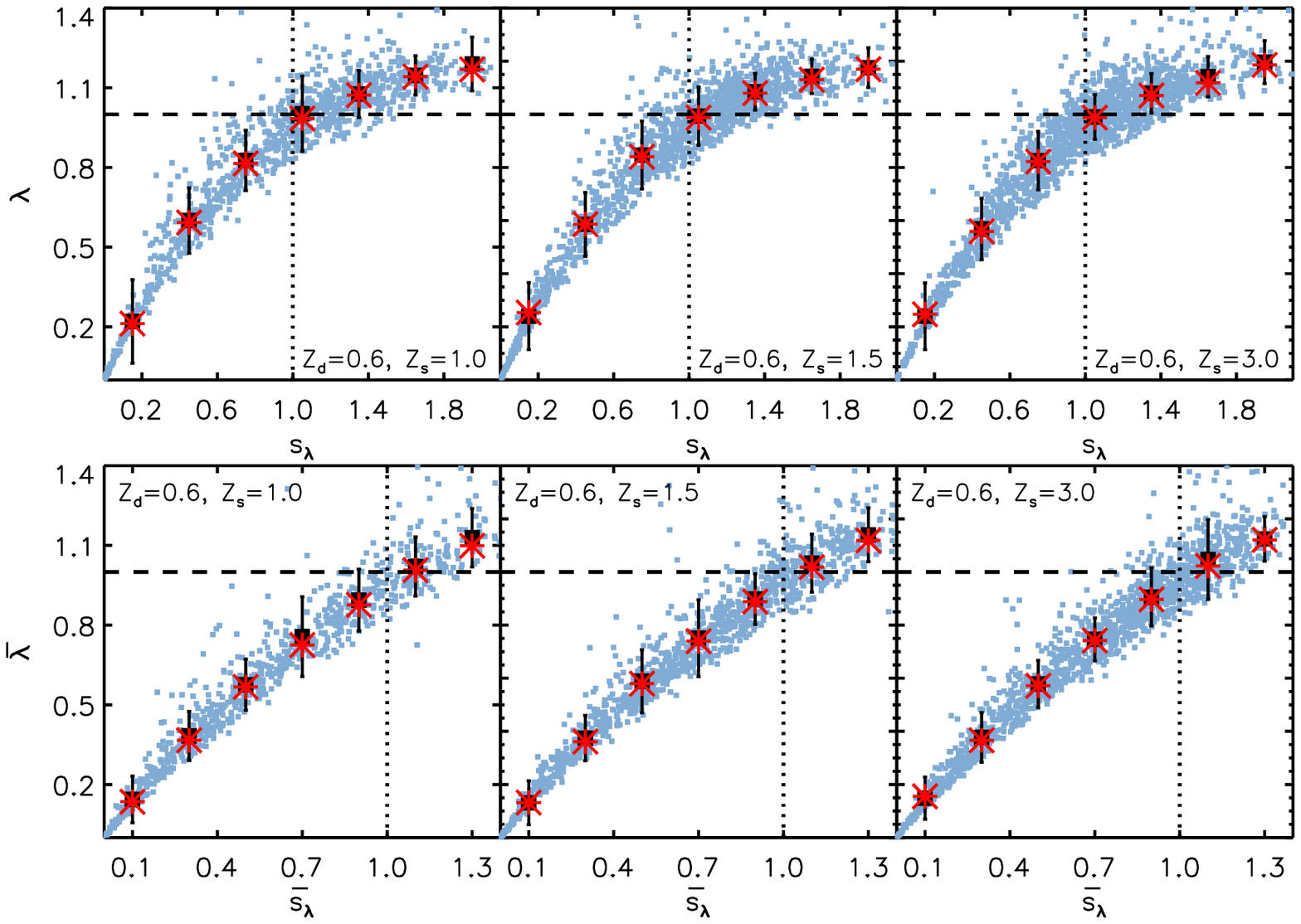}
\caption{Distributions of $\lambda$ versus $s_{\lambda}$ and
  $\bar{\lambda}$ versus $\bar{s}_{\lambda}$ for all six samples with
  $z_{\rm d}$ and $z_{\rm s}$ indicated in the panels. On top of the
  scattered data (blue dots) the black squares (and red stars) with
  the error bars indicate the mean (and median) and its standard
  deviation within each bin. The dashed lines indicate where
  $\lambda=1$ and where $\bar{\lambda}=1$; while the dotted lines
  indicate where $s_{\lambda}=1$ and where $\bar{s}_{\lambda}=1$.}
\label{fig:LambdaDistribution2}
\end{figure*}

\twocolumn
\bibliographystyle{mn2e}
\bibliography{ms_xudd}
\label{lastpage}

\end{document}